# Optical characteristics and capabilities of the successive versions of Meudon and Haute Provence Hα heliographs (1954-2004)

J.-M. Malherbe, Observatoire de Paris, PSL Research University, CNRS, LESIA, Meudon, France

Email : Jean-Marie.Malherbe@obspm.fr

ORCID id : https://orcid.org/0000-0002-4180-3729

Date : 13 April 2023

## ABSTRACT

Hα heliographs are imaging instruments designed to produce monochromatic images of the solar chromosphere at fast cadence (60 s or less). They are designed to monitor efficiently dynamic phenomena of solar activity, such as flares or material ejections. Meudon and Haute Provence observatories started systematic observations in the frame of the International Geophysical Year (1957) with Lyot filters. This technology evolved several times until 1985 with tunable filters allowing to observe alternatively the line wings and core (variable wavelength). More than 6 million images were produced during 50 years, mostly on 35 mm films (catalogs are available on-line). We present in this paper the optical characteristics and the capabilities of the successive versions of the Hα heliographs in operation between 1954 and 2004, and describe briefly the new heliograph (MeteoSpace) which will be commissioned in 2023 at Calern observatory.

## INTRODUCTION

Spectroheliographs (Deslandres, 1910; d'Azambuja, 1930) were intensively employed in the first half of the twentieth century to study the evolution of the solar atmosphere (note 1) and structures (note 2), during four or five cycles of activity, in various chromospheric lines such as Hα and CaII K. However, spectroheliographs are spectroscopic instruments; it takes time for the entrance slit to scan the solar surface, so that the temporal cadence of observations was rather slow; the use of photographic plates (13 x 18 cm²) also complicated the task. This is the reason why instruments with monochromatic filters and 35 mm films appeared in the fifties; they are more convenient for the survey of fast events of solar activity, such as flares and material ejections, which are highly dynamic phenomena and evolve on short time scales (several minutes). Birefringent filters were invented by Lyot in the thirties (Lyot, 1944). The transmittance is centred on a Fraunhofer line, but in comparison to Fabry-Pérot filters, they are more suitable to isolate a narrow bandpass, because their wings are lower and avoid contrast reduction by parasitic photospheric light; this property does not play much when observing prominences at the limb, but it becomes important for chromospheric structures seen on the disk (such as filaments), which are illuminated below by the photospheric continuum. Also, 35 mm films can record 1850 frames, so that a complete film, with three images per minute, has a duration of 10 hours, which is well adapted to record one day of solar activity. Meudon and Haute Provence (OHP) heliographs started systematic surveys with Lyot filters in the context of the International Geophysical Year (IGY 1957, see d'Azambuja, 1959). Several generations of Lyot filters, including tunable devices, were used in Meudon until 2004. Fast tunable filters are based on rotating birefringent plates and can observe the line centre and wings of Hα within a few seconds; this feature is a great advantage for observations of dynamic events which exhibit mass motions revealed by the presence of Dopplershifts (note 3).

Section 1 presents the principles of the monochromatic birefringent Lyot filters, while Section 2 describes the initial versions of the Hα heliographs designed for Meudon and Haute Provence observatories. Section 3 is dedicated to the Meudon tunable (variable wavelength) filters, both the 1965 and 1985 versions, and Section 4 summarizes systematic observations of solar activity performed between 1956 and 2004. At last, Section 5 presents the MeteoSpace project for a new heliograph, that will be commissioned in the course of 2023 at the Calern observatory.

# 1 – THE PRINCIPLE OF MONOCHROMATIC BIREFRINGENT LYOT FILTERS

The LYOT filter is a birefringent, polarising and monochromatic filter, with a narrow bandwidth, usually 0.5 Å for Hα, but it may be as low as 0.05 Å for exceptional instruments, such as the NFI filter onboard the Hinode/JAXA/NASA satellite (launched in 2006). The basic principle is the interference of the ordinary and extraordinary rays at the exit of a birefringent crystal, such as spath or quartz. However, spath presents more than 15 times the birefringence of quartz, as shown by figures 1 and 2. This property allows the construction of more compact filters, because the phase shift (rd) between ordinary and extraordinary rays is equal to:

$$\delta = (2\pi/\lambda)\ \Delta n\ e$$ where $\Delta n = |n_o - n_e|$ is the difference of the refractive index and e the thickness of the crystal.

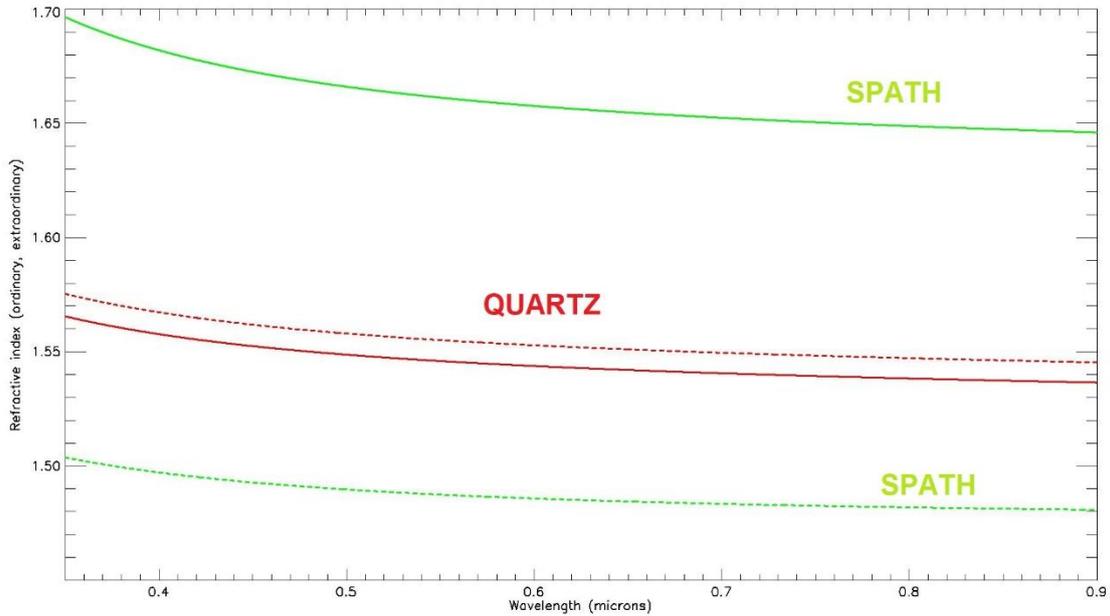

**Figure 1** : *Refractive index of the ordinary (solid line) and extraordinary rays (dashed line) for spath (CaCO₃) and quartz (SiO₂), as a function of wavelength (micrometres). The difference $\Delta n = |n_o - n_e|$ is about 0.17 for spath and only 0.009 for quartz. Courtesy Paris observatory.*

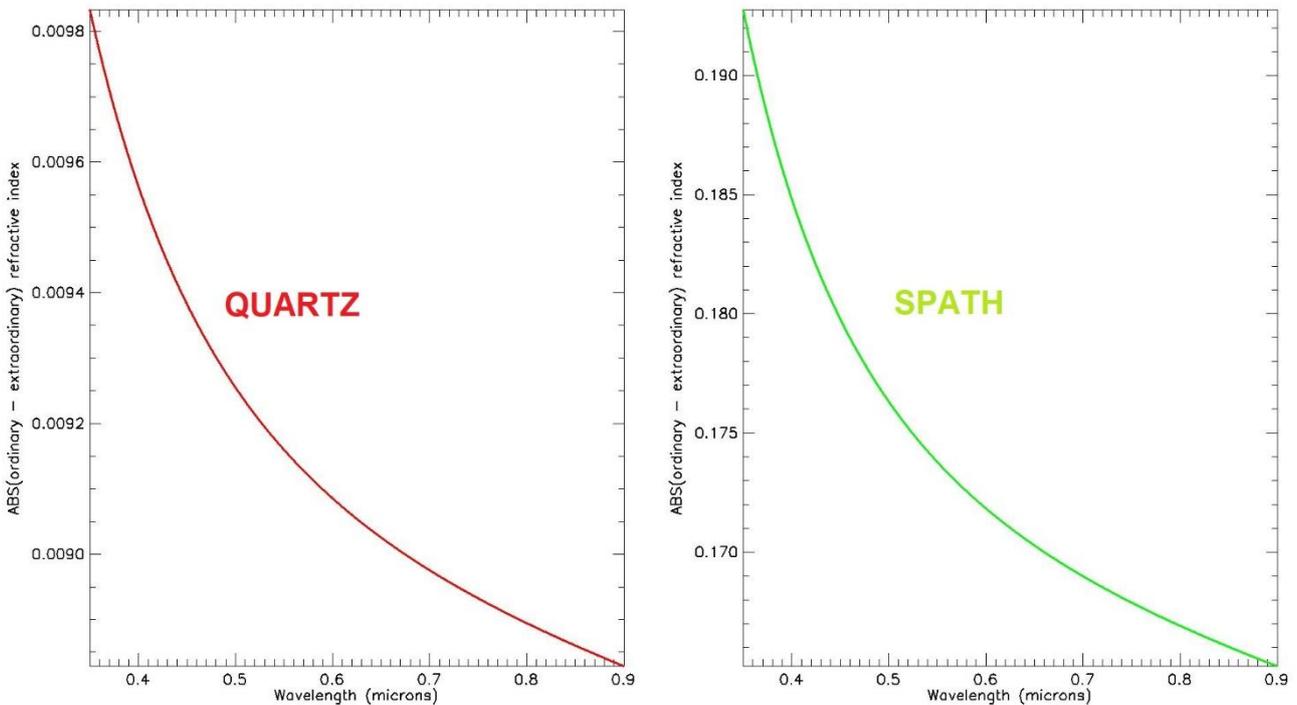

**Figure 2** : *Absolute value of the difference of the ordinary and extraordinary refractive index for spath and quartz, as a function of wavelength (micrometres). The difference $\Delta n = |n_o - n_e|$ is about 0.17 for spath and only 0.009 for quartz. Courtesy Paris observatory.*

The Lyot filter has in general many stages (5 to 8) composed each of a birefringent crystal between two linear polarizers. The Fast (F) and Slow (S) axis of the crystal are orthogonal to the beam direction and have an azimuth of 45° with respect to the direction of the polarizers. The Fast axis corresponds to the smallest refractive index, while the Slow axis to the highest. For spath crystals, $n_o > n_e$, while it is the contrary for quartz. Each stage delivers a channeled spectrum with periodic maxima and null minima. The thickness of stage n is twice the one of stage n-1 (e, 2 e, 4 e, … $2^{n-1}$ e). The maxima of stage n occur at the positions of both the maxima and minima of stage n-1, providing when n = 6 (or more) narrow transmission peaks such as those of figure 3. The useful peak is finally isolated by a coloured (or interference) filter.

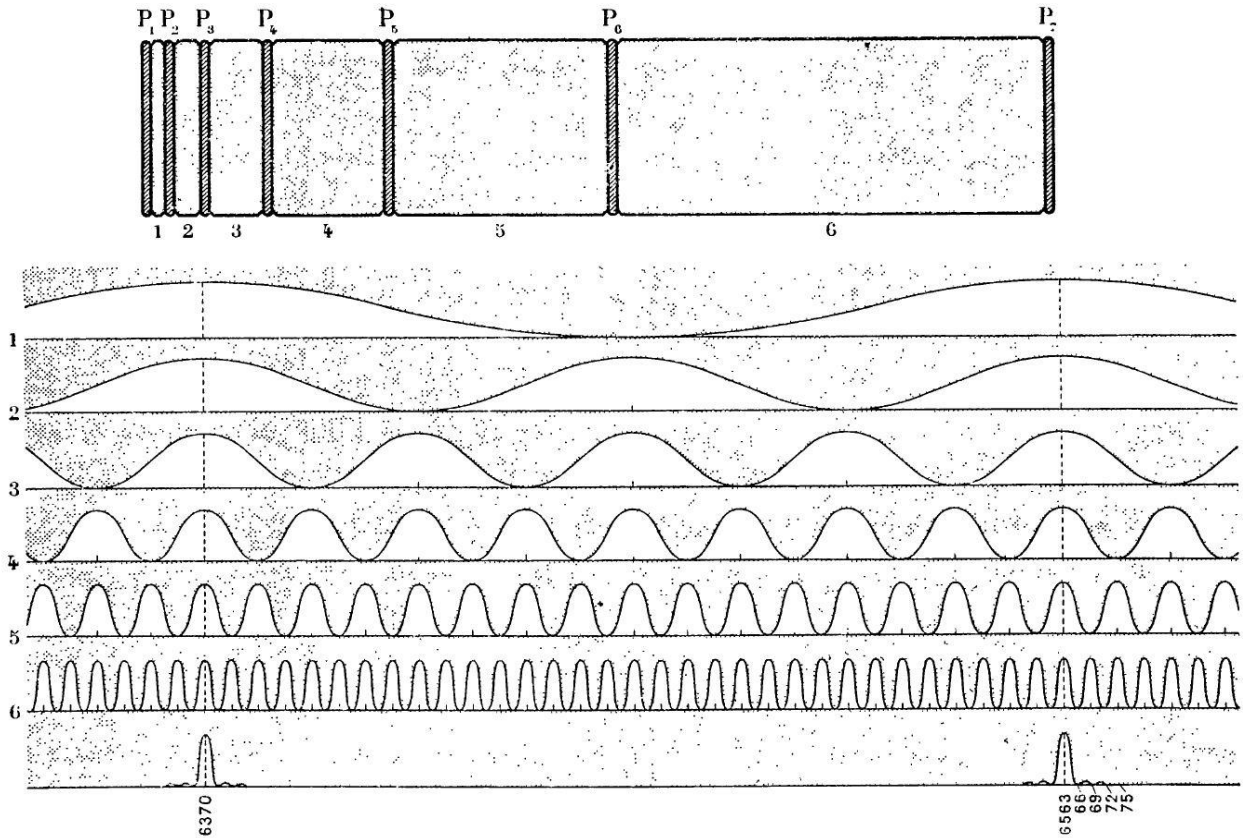

**Figure 3** : *Six stages Lyot filter made of quartz between polarizers (6 crystals, 7 polarizers). Fast and Slow axis are at 45° from the polarizing direction. The transmission of each stage (channelled spectrum) is displayed and their multiplication provides the overall transmission reported at the bottom (in abscissa: the wavelength). This filter allowed to observe the iron coronal line (6374 Å) and Hα (6563 Å). After Lyot (1944).*

The filter of figure 3 has n = 6 quartz stages of thickness e = 2.22 mm, 2 e = 4.44 mm, 4 e = 8.9 mm, 8 e = 17.8 mm, 16 e = 35.5 mm and $2^{n-1}$ e = 32 e = 71.0 mm, giving the total thickness of $(2^n - 1)$ e = 63 e = 140 mm (without the polarizers). Such a filter has a bandwidth of 3.0 Å FWHM (details below). One sees that it would be impossible to design very selective filters (FWHM < 1.0 Å), because of the thickness of quartz plates. For that reason, composite filters were introduced by Lyot, with some plates in quartz and the other ones in spath. For example, it is possible to add to this 6 quartz stages filter, two other spath stages of thickness 7.5 mm and 15 mm to get a 0.75 Å FWHM transmission.

Lyot filters are temperature regulated (value in the range 30°C - 45°C) in order to stabilize the optical thickness, otherwise the central wavelength should fluctuate a lot. The precision of the controller must be about 0.1°C with active electric heating (Joule effect) and passive cooling (conduction). The optics is embedded inside electric resistors (40 Ω typical, 12 V) and is thermally isolated by cork (or similar material). A temperature sensor is glued directly upon the birefringent plates; another one measures the ambient temperature. The heating power is moderate (about 5 W) and pulsed.

The initial filters were designed to observe the core of spectral lines. For narrow band filters (FWHM < 1.0 Å), it is usefull to move the transmission peak to explore the wings (for instance to estimate Dopplershifts), but one cannot wait for temperature adjustment, because the thermal inerty makes the process too slow. Lyot proposed a much faster opto-mechanical solution with rotating birefringent plates associated to a quarter-wave plate (details below). In order to displace precisely the waveband, each stage rotates; for plates of thickness e, 2 e, 4 e, … $2^{n-1}$ e, the position angle is α, 3 α, 7 α, … $(2^n-1)$ α, so that the rotation angle between stage n-1 and stage n is $2^{n-1}$ α. This implies a complex mechanics rotating in a bath of optical oil.

## 1 – a - An elementary Lyot stage

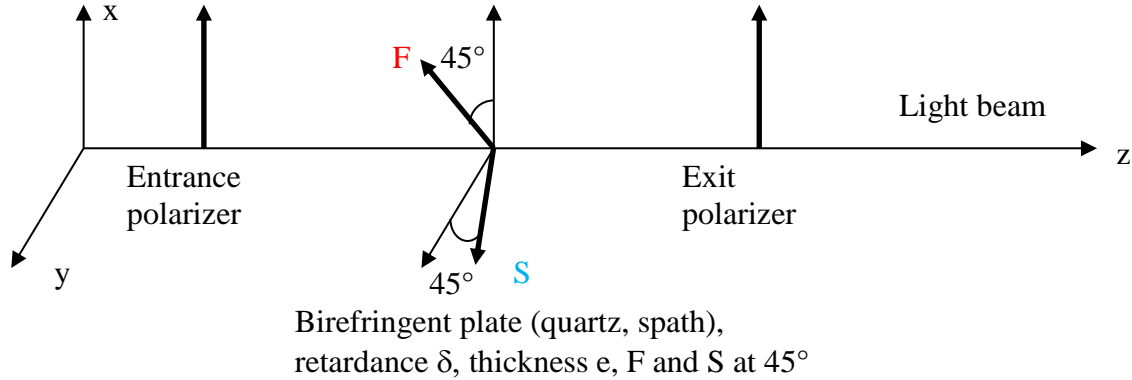

**Figure 4** : *An elementary Lyot stage between two parallel polarizers. Courtesy Paris observatory.*

The birefringent plate is located between two parallel polarizers; the Fast (F) and Slow (S) axis of the crystal make an angle of 45° with respect to the polarizers (figure 4). The plate of thickness e introduces a phase lag between the projections of the X-axis vibration (entrance polarizer) along the F and S axis, which is equal to δ = (2π/λ) Δn e, where Δn = |$n_o$ − $n_e$| is the difference of the ordinary and extraordinary refractive index (for spath, $n_o$ = 1.658 and $n_e$ = 1.486). The exit polarizer recombines the vibrations along F and S. The emerging intensity I is related to the incident intensity $I_0$ by the relationship:

I = $I_0$ $\cos^2$ (δ/2), where  δ = (2π/λ) Δn e

This is a channeled spectrum, as those of figure 3.

## 1 – b – A study of a basic Lyot filter

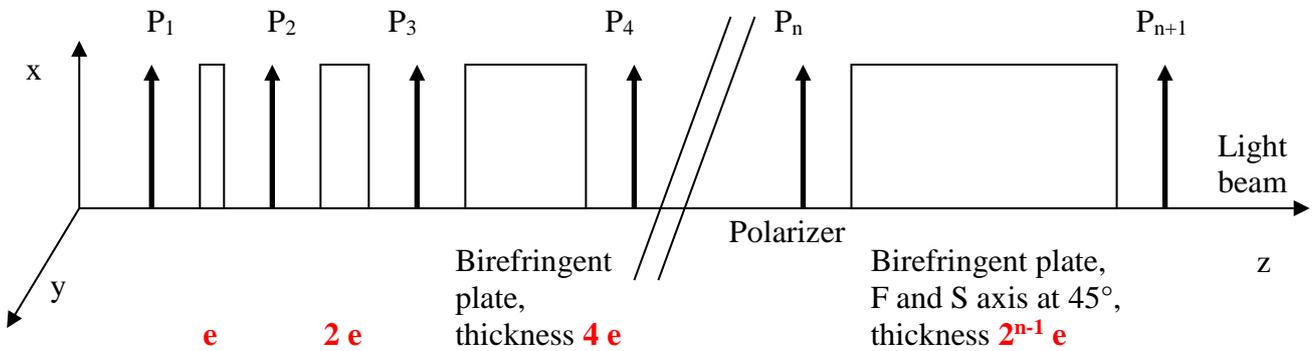

**Figure 5** : *A basic Lyot filter with n stages and (n+1) polarizers. Courtesy Paris observatory.*

Let us consider a n stages filter (figure 5). It is made of n elementary stages (figure 4) with n+1 polarizers (X-axis), denoted $P_i$, and n birefringent plates with Fast axis (F) and Slow axis (S) at 45° from the direction of the polarizers. The thickness of the plates varies in power of 2, so that the successive thicknesses are e, 2 e, 4 e, 8 e, 16 e, … and $2^{n-1}$ e. The emerging intensity I is related to the incident intensity $I_0$ by the relationship:

I = $I_0$ $\cos^2$ (δ/2) $\cos^2$ (2(δ/2)) $\cos^2$ (4(δ/2))......$\cos^2$ ($2^{n-1}$(δ/2)),   where δ = (2π/λ) Δn e

This formula can be simplified, and finally we get:

I = $I_0$ [ sin ( $2^n$ (δ / 2) ) / ( $2^n$ sin ( δ/2 ) ) ]$^2$,   where δ = (2π/λ) Δn e

It provides the channelled spectra similar to those of figures 6 and 7.

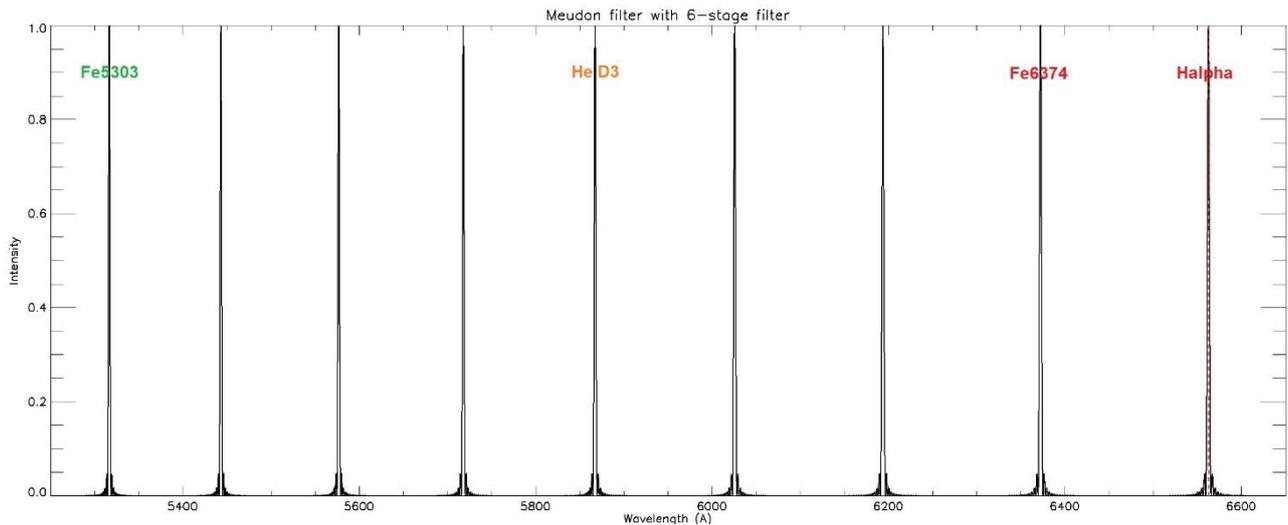

**Figure 6** : *Channeled spectrum of a Lyot filter with n = 6 quartz stages, e = 2.22 mm (thickness of plates: e, 2 e, 4 e, 8 e, 16 e, 32 e), 3.0 Å FWHM. Wavelength in abscissa (Å). This filter allowed observations of the chromospheric Hα and He D3 lines, as well as two forbidden coronal lines (the red and green lines of highly ionized iron). A coloured filter can isolate the peak of interest. Courtesy Paris observatory.*

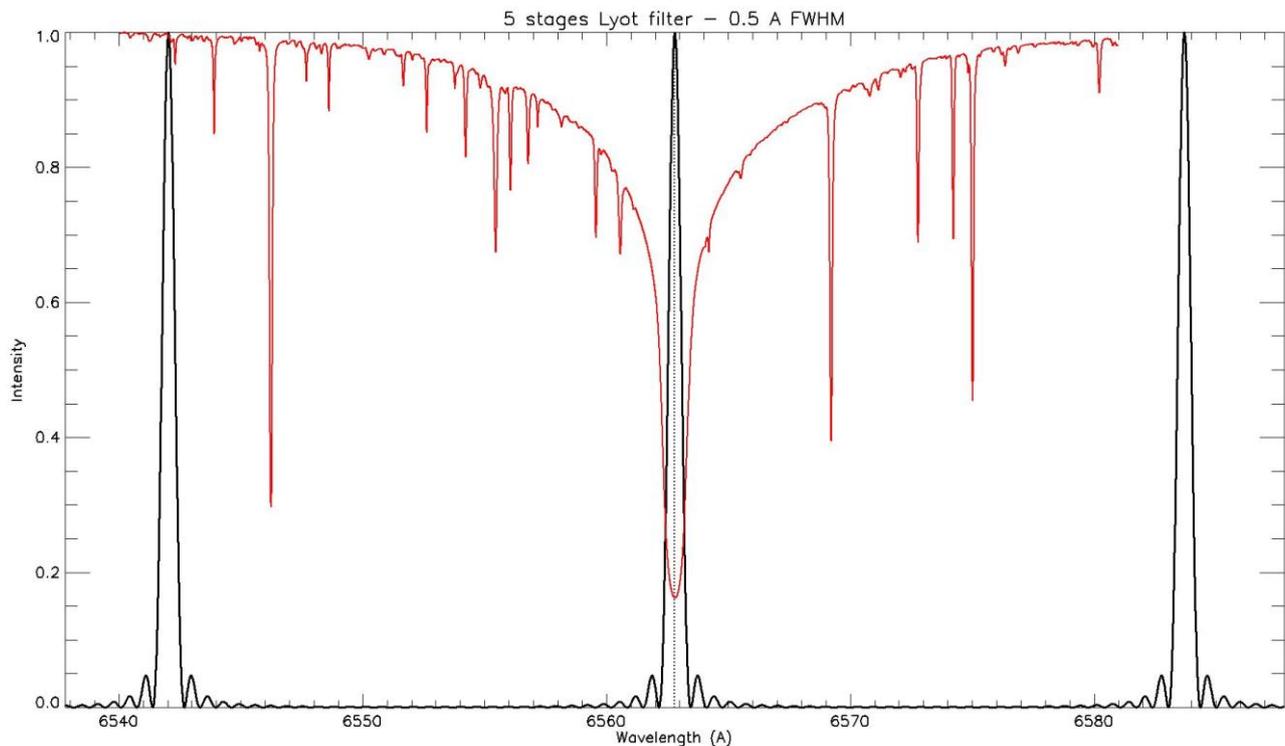

**Figure 7** : *Channeled spectrum of a LYOT filter with n = 5 spath stages, designed for the Hα line, e = 1.2 mm (thickness of plates e, 2 e, 4 e, 8 e, 16 e), 0.5 Å FWHM. Wavelength in abscissa (Å). The Hα line is in red. The peak of interest must be isolated using a narrow interference filter. Courtesy Paris observatory.*

The wavelengths $λ_m$ of the main peaks (maxima of transmission) are:

$λ_m = Δn\ e\ /\ k$   (k integer, the interference order, peaks are not equidistant in wavelength, <span style="color:cyan">figure 6</span>)

The wavelength distance Dλ between two successive peaks is:

$Dλ = Δn\ e\ /\ k^2\ = λ_m^2\ /\ Δn\ e$        (not a constant)

For a quartz filter (Δn = 0.009), the distance between two successive peaks is large (<span style="color:cyan">figure 6</span>) and the peak of interest can be isolated with coloured filters; but for a spath filter (Δn = 0.17), and comparable thickness e, this distance is small (<span style="color:cyan">figure 7</span>), so that a narrow interference filter is needed to select the peak of interest.

Wavelength locations of zéro transmission:

$$\lambda = 2^n \, \Delta n \, e \, / \, p \quad \text{(p integer)}$$

The full width at halft maximum $\Delta\lambda$ of main peaks (FWHM) is approximately equal to the distance between two successive zeros given by $|\Delta\delta| = 2\,\pi \, / \, 2^n = \Delta\lambda \, (2\pi/\lambda_m^2) \, \Delta n \, e$, from which we derive the result :

$$\Delta\lambda = \lambda_m^2 \, / \, (2^n \, \Delta n \, e)$$

The finess of the filter is the ratio: $D\lambda \, / \, \Delta\lambda = 2^n$   (depends only of the number of stages)

This relation means that the finess doubles when a stage is added to a Lyot filter.

## 1 – c – A wide field Lyot filter

For observations of the full solar disk (an extended object of mean angular diameter 32'), the design of standard Lyot filters exhibits variations of transmission from the centre to the limb. Lyot showed that is effect can be corrected. The principle consists of replacing the birefringent plate of thickness e (figure 4) by two crossed birefringent plates of thickness e/2 (figure 8), separated by a half wave plate oriented at 45° of the Fast (F) and the Slow (S) axis of the two birefringent plates. The half wave plate is essential: it acts as a polarization rotator of $\pi/2$, so that the F axis of plate 1 will rotate by $\pi/2$ to correspond between the F axis of plate 2 (the same mechanism operates for the S axis). With the half wave plate, the retardances of the two birefringent plates of thickness e/2 become additive instead of being subtractive, so that the global phase lag of the system is still the one of the elementary stage of figure 4:

$$\delta = (2\pi/\lambda) \; [ \; n_o \, (e/2) + n_o \, (e/2) - n_e \, (e/2) - n_e \, (e/2) \; ] = (2\pi/\lambda) \, \Delta n \, e$$

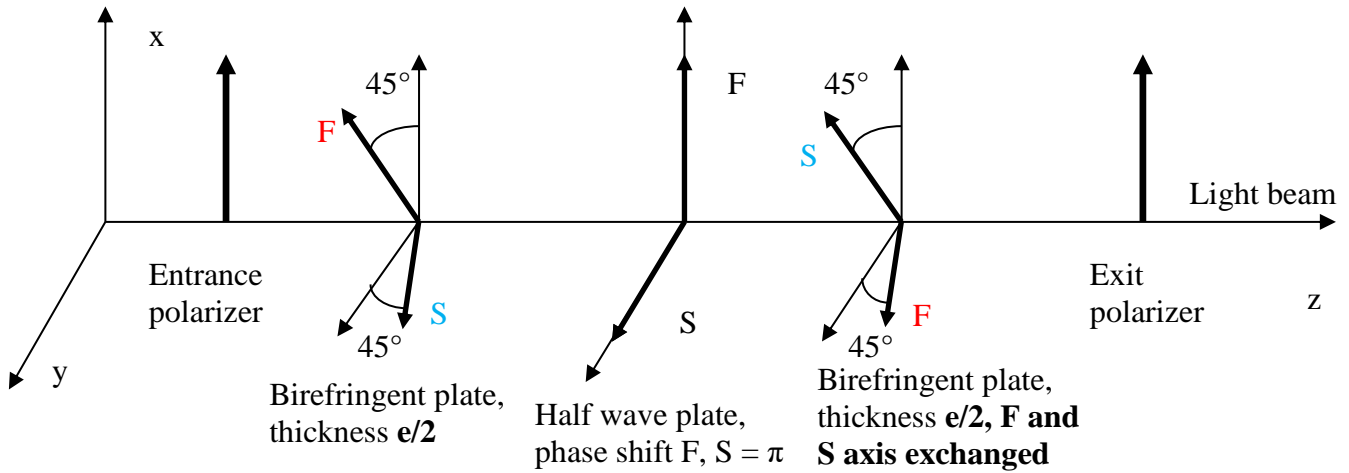

**Figure 8** : *An elementary stage of wide field Lyot filters. The birefringent plate of thickness e is replaced by two crossed birefringent plates of thickness e/2. A half wave plate is incorporated between, in order to rotate the polarization ($\pi/2$ angle), so that the F axis of plate 1 will correspond to the F axis of plate 2 (and the same for S axis). Courtesy Paris observatory.*

We show below that the crossing of the two birefringent plates of thickness e/2 considerably reduces the angular dependence of the transmission function; this is the reason why many monochromatic heliographs designed for full solar disk observations use wide field Lyot filters.

For the classical design with only one birefringent plate of thickness e, we introduce i, the incidence angle of the beam (a small angle in paraxial optics) and $\theta$, the azimuth of the beam ($0 < \theta < 2\pi$) with respect to the optical axis. The phase lag $\delta(i, \theta)$ for angles i and $\theta$ between the ordinary and extraordinary rays is now:

$$\delta(i, \theta) = \delta(0) \; [ \; 1 - i^2 \, (\cos^2 \theta - (n_o/n_e) \, \sin^2 \theta) \, / \, (2 \, n_o^2) \; ]$$

where $\delta(0) = (2\pi/\lambda) \, \Delta n \, e$ is the phase lag for null incidence.
From this formula, we can derive the corresponding wavelength shift :

$\Delta\lambda(i, \theta) / \lambda = - i^2 (\cos^2 \theta - (n_o/n_e) \sin^2 \theta) / (2 n_o^2)$

This last relationship indicates that :

- $\Delta\lambda(i, \theta)$ is of second order with respect to the incidence angle i
- with $\theta = 0$,   $\Delta\lambda(i) = \lambda [ - i^2 / (2 n_o^2) ]$   (typically $\Delta\lambda(i) = \lambda [ - 0.18 i^2 ]$ for spath)
- with $\theta = \pi/2$, $\Delta\lambda(i) = \lambda [ + i^2 / (2 n_o n_e) ]$   (typically $\Delta\lambda(i) = \lambda [ + 0.20 i^2 ]$ for spath)
- according to azimuth $\theta$, the wavelength shift is either towards the red ($\theta = \pi/2$) or the blue ($\theta = 0$)

Now, for the wide field device (index "w" below) with two crossed birefringent plates of thickness e/2, the formula above, providing the phase lag, applies two times, so that :

$\delta_w(i, \theta) = \frac{1}{2} ( \delta(i, \theta) + \delta(i, \theta + \pi/2) ) = \delta(0) [ 1 - i^2 (n_e - n_o) / (4 n_o^2 n_e) ]$

The phase lag of the wide field design is no more $\theta$ dependant. The corresponding wavelength shift is now :

$\Delta\lambda_w(i) / \lambda = - i^2 (n_e - n_o) / (4 n_o^2 n_e)$

Typically $\Delta\lambda(i) = \lambda [ + 0.01 i^2 ]$ for spath, one sees that there is a redshift, but it is about 20 times reduced in comparison to the classical design (0.07 Å for Hα at F/15 instead of 1.4 Å, value not acceptable).

## 1 – d – A tunable or variable wavelength Lyot filter

In order to evaluate qualitavely mass motions, which are the major phenomena of solar activity (flares and ejections), it is necessary to determine Dopplershifts. One knows that upward motions are at the origin of blueshifts of spectral lines, while downward motions generate redshifts. The observation of line wings allows to determine surely the direction of the motion (up or down), and delivers an approximate value of the line of sight velocity (1 Å = 45 km/s in Hα). For that reason, the design of tunable (or variable wavelength) Lyot filters was a true challenge for the survey of solar activity. This can be done by temperature variation, but it is an extremely slow process because of the thermal inerty of the filter. Lyot proposed a considerably faster method, an opto-mechanical solution based on the rotation of the filter stages. This means that each individual stage must be included in a rotating mount, contrarily to the initial filter where the successive stages were glued together.

Let us consider the elementary stage of figure 9. The entrance polarizer is now inclined in the (X, Y) plane with angle α. It is followed by a quarter wave plate, the birefringent plate of thickness e (which can also be the wide field device of figure 8 with two crossed birefringent plates of thickness e/2 separated by the half wave plate) and the exit polarizer. A similar entrance polarizer in the X-direction followed by the three following optical elements (quarter wave plate, birefringent plate, exit polarizer), all rotated by angle α, is strictly equivalent (and used in practice).

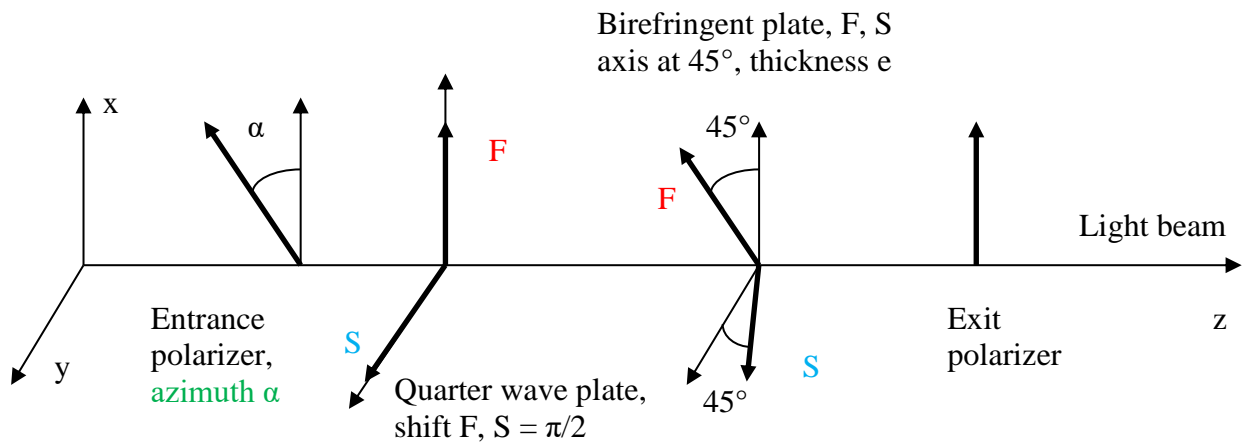

**Figure 9** : *A tunable elementary stage of Lyot filter. The quarter wave plate transforms the linear vibration (polarizer) into an elliptic one, and introduces a supplementary and variable phase shift α between the ordinary and extraordinary vibrations along the F and S axis of the birefringent plate. Courtesy Paris observatory.*

One shows easily that the emerging intensity I is related to the incident intensity $I_0$ by the relationship:

$I = I_0 \cos^2(\delta/2 - \alpha)$  where $\delta = (2\pi/\lambda) \, \Delta n \, e$, $\Delta n = |n_o - n_e|$

The effect of the quarter wave plate is just to introduce a phase shift $\alpha$ in the classical formula of the elementary stage, for which $I = I_0 \cos^2(\delta/2)$. The result is again a channeled spectrum, but the peaks can be displaced quickly just by rotating the Lyot stage, which is composed of the quarter wave plate, the birefringent plate and the exit polarizer (the entrance polarizer is fixed). Let us know consider a n stages filter (figure 10) :

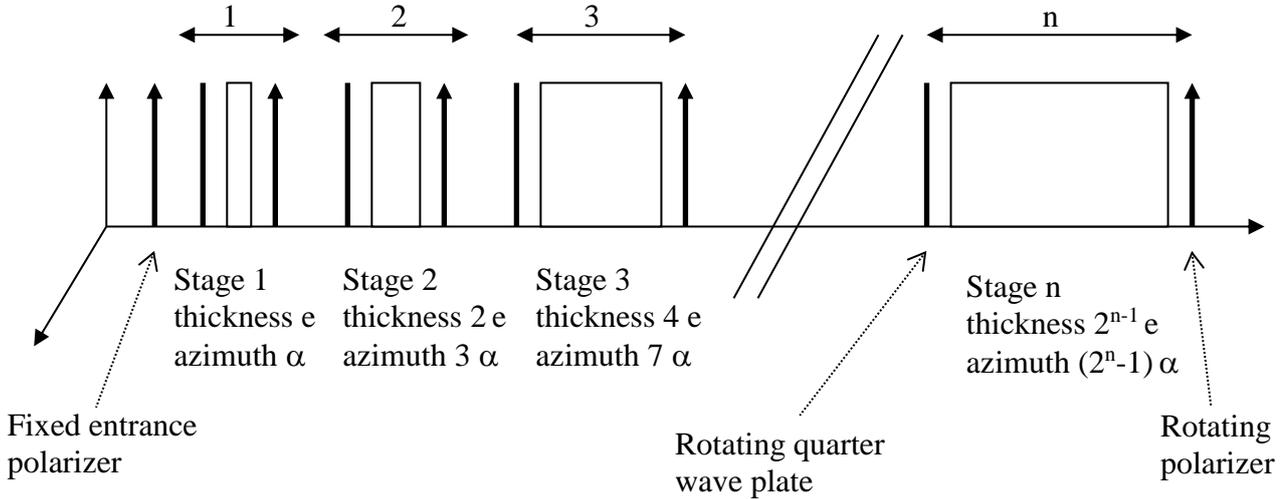

**Figure 10** : *Tunable Lyot filter with n rotating stages, each one composed of a quarter wave plate, a biregringent plate and a polarizer. The entrance polarizer is fixed. Courtesy Paris observatory.*

In this optical combination of n successive stages :
- the entrance polarizer is fixed
- stage 1 (thickness e) is rotated by angle $\alpha$
- stage 2 (thickness  2 e) is rotated by angle $2\alpha$ with respect to stage 1, or $3\alpha$ with respect to the entrance polarizer
- stage 3 (thickness  4 e) is rotated by angle $4\alpha$ with respect to stage 2, or $7\alpha$ with respect to the entrance polarizer
- stage n (thickness $2^{n-1}$ e) is rotated by angle $2^{n-1}\alpha$ with respect to stage n-1, or $(2^n-1)\alpha$ with respect to the entrance polarizer

A n stages filter has the following emergent intensity :

$I = I_0 \cos^2(\delta/2-\alpha) \cos^2(2(\delta/2-\alpha)) \cos^2(4(\delta/2-\alpha))......\cos^2(2^{n-1}(\delta/2-\alpha))$  where $\delta = (2\pi/\lambda) \, \Delta n \, e$

This formula can be simplified, and finally we get :

$I = I_0 \left[ \sin ( 2^n(\delta/2 - \alpha) ) / ( 2^n \sin(\delta/2 - \alpha) ) \right]^2$  where $\delta = (2\pi/\lambda) \, \Delta n \, e$

When $\alpha$ varies, the maximum displacement is the distance between two successive peaks.

### 1 – e – Interlacing two Lyot filters

Interlacing two Lyot filters is an interesting method to reduce the amplitude of secondary peaks around the primary peaks. Let us consider a n stages filter of thickness e, 2 e, 4 e … $2^{n-1}$ e, and a m stages filter of thickness d, 2 d, 4 d … $2^{m-1}$ d. The emerging intensity is given by the product:

$I = I_0 \left[ \sin ( 2^n(\delta/2 - \alpha) ) / ( 2^n \sin(\delta/2 - \alpha) ) \right]^2 \; \left[ \sin ( 2^m(\gamma/2 - \beta) ) / ( 2^m \sin(\gamma/2 - \beta) ) \right]^2$

where $\delta = (2\pi/\lambda) \, \Delta n \, e$  and  $\gamma = (2\pi/\lambda) \, \Delta n \, d$

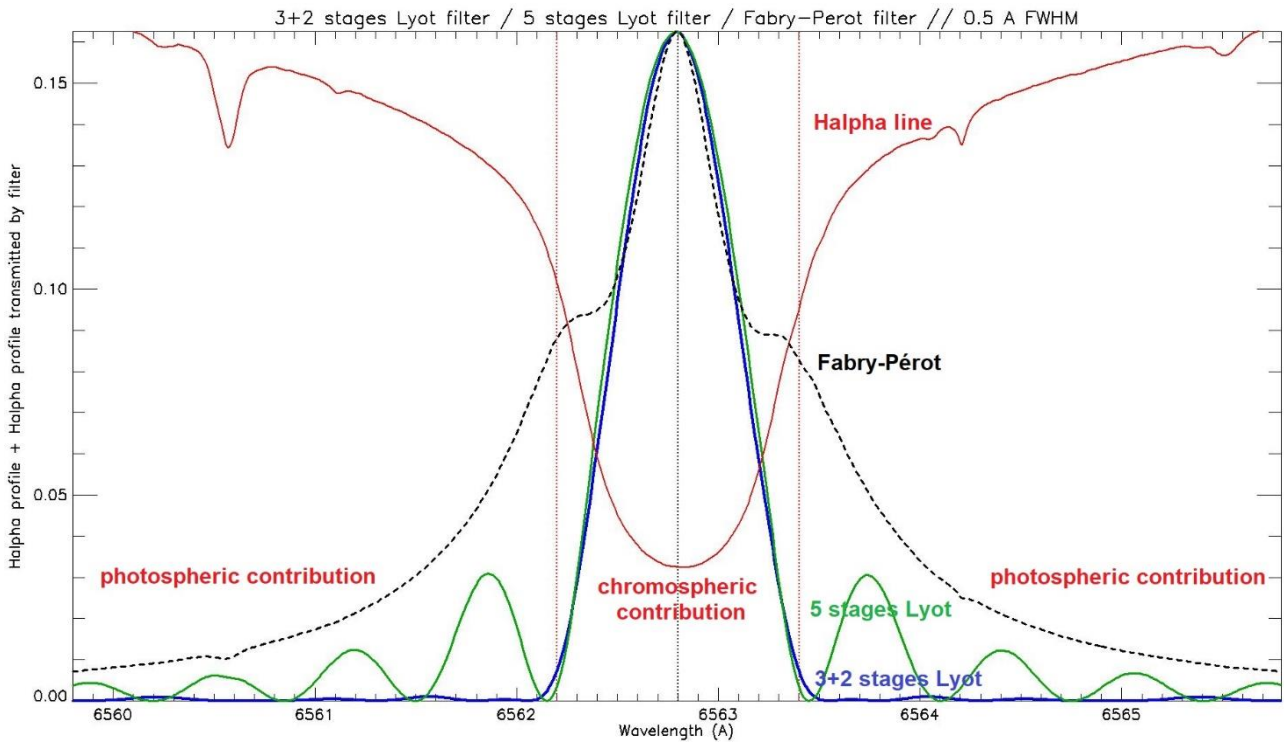

**Figure 11** : *Hα line profile transmitted by an interlaced LYOT filter (*<span style="color:blue">blue</span>* curve) with 3 stages of thickness e, 2 e, 4 e (e = 4.2 mm) and 2 stages of thickness d, 2 d with d = 3/2 e. The FWHM is 0.54 Å. For comparison, the line transmitted by a classical 5 stages Lyot filter (*<span style="color:green">green</span>* curve) with same FWHM, showing parasitic secondary peaks. The Hα line profile is in *<span style="color:red">red</span>*. The dashed line provides the Hα line transmission by a Lorentzian filter (Fabry-Pérot) of similar FWHM (wings are much broader). Courtesy Paris Observatory.*

The values of d and β are chosen such that d / e = β / α = k (real number). When k is a rational fraction, the primary peaks of both filters are superimposed and the secondary peaks of the first filter roughly correspond to zeros of the second one, so that they are cut; this is a great advantage for observations of chromospheric lines, because it eliminates parasitic light coming out from the photosphere through the line wings. As a consequence, the contrast of structures, such as filaments, is enhanced. Figure 11 displays the Hα line profile together with its multiplication by the transmittance of the filter. With k = 3/2, figure 11 shows that the interlaced filter (<span style="color:blue">blue</span> curve) has no secondary peaks in comparison to the classical design (<span style="color:green">green</span> curve). On the contrary, Fabry-Pérot filters (Lorentzian shape) generate extended wings, which decrease the contrast of chromospheric structures.

### 1 − f − Liquid Crystals Variable Retarders (LCVR) in Lyot filters

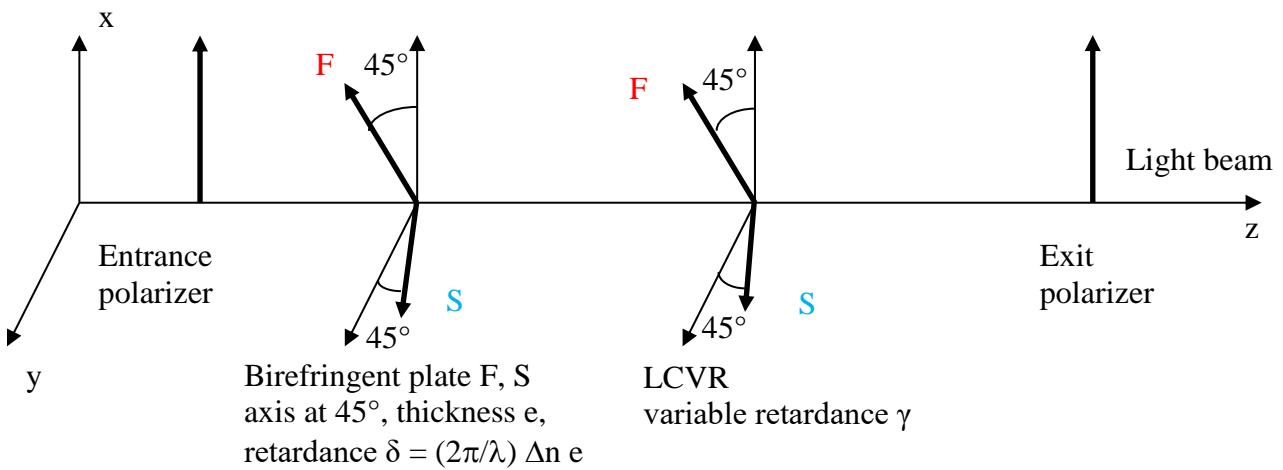

**Figure 12** : *A tunable elementary stage of Lyot filter. The LCVR introduces a supplementary and variable phase shift γ between the ordinary and extraordinary rays. Courtesy Paris observatory.*

Recently, Liquid Crystal Variable Retarders (LCVR) appeared, they provide a variable retardance γ (from 0 to 2π). Fast (F) and Slow (S) axis of the LCVR and the birefringent plate are parallel (figure 12), so that the total retardance becomes (δ + γ) and the emerging intensity I est related to the incident intensity $I_0$ by the formula :

$$I = I_0 \cos^2(\ (\delta+\gamma)\ /\ 2\ ) \quad \text{where } \delta = (2\pi/\lambda)\ \Delta n\ e, \ \ \Delta n = |n_o - n_e|$$

The transmission peaks move when the retardance γ of the LCVR electrically changes, under a varying electric field. From one stage to the next, the retardance must vary proportionally to the thickness, i.e. in power of 2. Unfortunately, we do not own such a device at Meudon.

## 2 – CAPABILITIES OF THE FIRST MEUDON AND OHP LYOT FILTERS

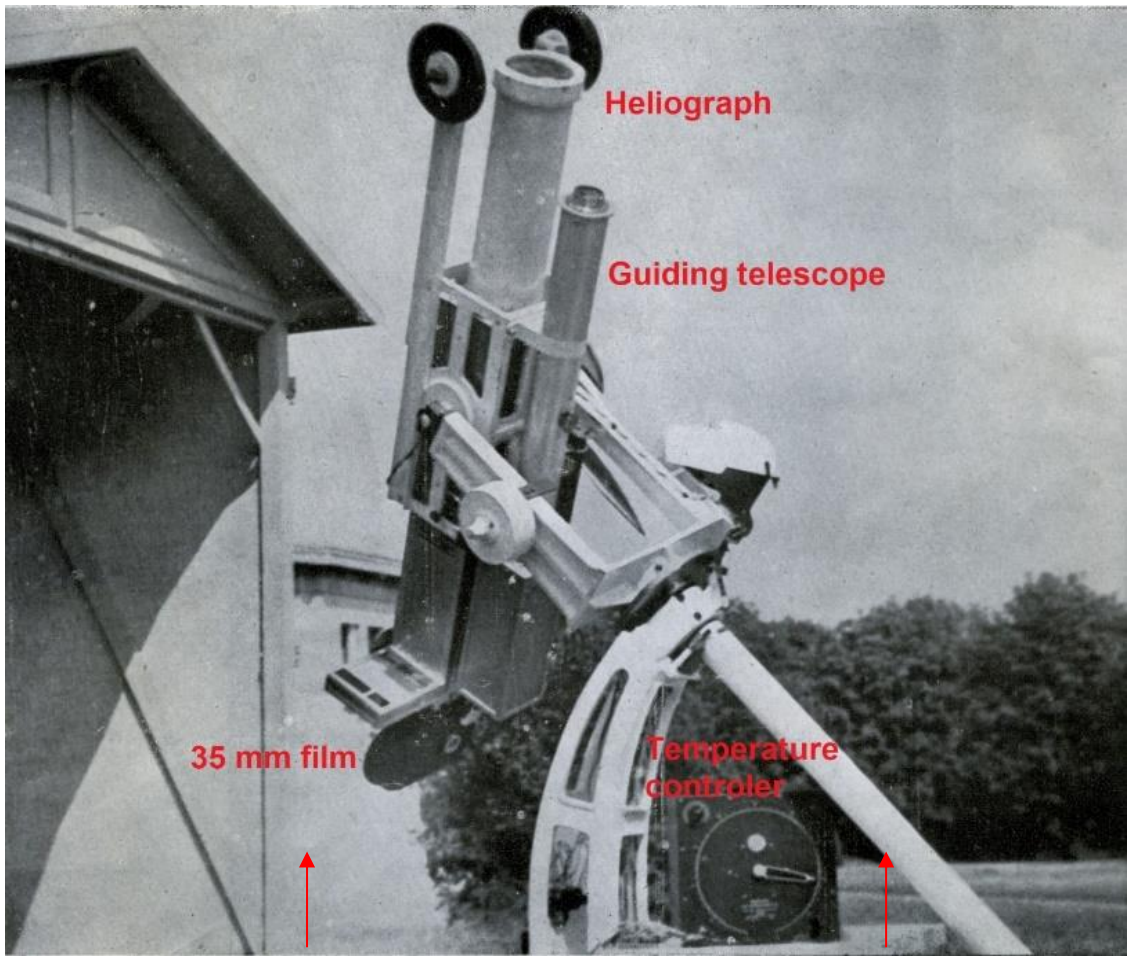

**Figure 13** : *The first heliograph of Meudon observatory (1954) and the guiding telescope. The 35 mm camera and the temperature controller of the filter are indicated. Courtesy Paris observatory. After Grenat & Laborde (1954).*

The first Meudon heliograph (figure 13) was decribed by Grenat & Laborde (1954). The focal length of the refractor was 1.40 m at F/10. A diverging lens allowed to magnify the solar image to 15 mm diameter, corresponding to an equivalent focal length of 1.60 m. The bandpass of the Lyot filter was 0.75 Å and it was centred in Hα. Systematic observations started in 1956 with 35 mm films of 45 metres each. The equatorial mount carried a guiding telescope with photoelectric cells using about 10% of the light.

The original 0.75 Å FWHM Lyot filter was duplicated at least 12 times by the "Optique et Precision de Levallois", a french company (OPL, figures 14 and 15), for many observatories world wide, in the frame of IGY1957 (d'Azambuja, 1959). A standard heliograph was manufactured by the SECASI company of Bordeaux (France), with this filter, and disseminated in six places such as observatories of McMath-Hulbert, Boulder (USA), Haute Provence (France), Mitaka (Japan), Purple Mountain (China), and Cape of Good Hope (Africa).

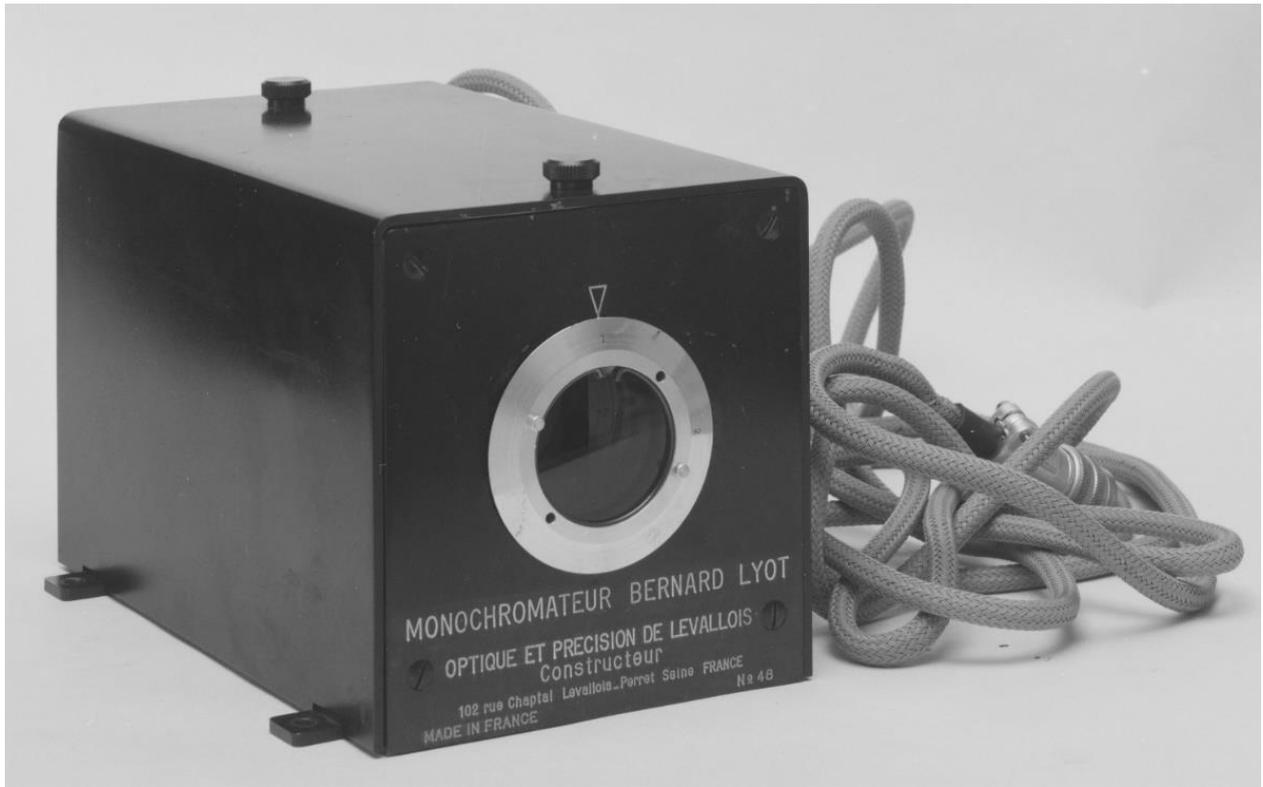

**Figure 14** : *The Lyot filter (0.75 Å FWHM) manufactured by the OPL company. Courtesy Paris observatory.*

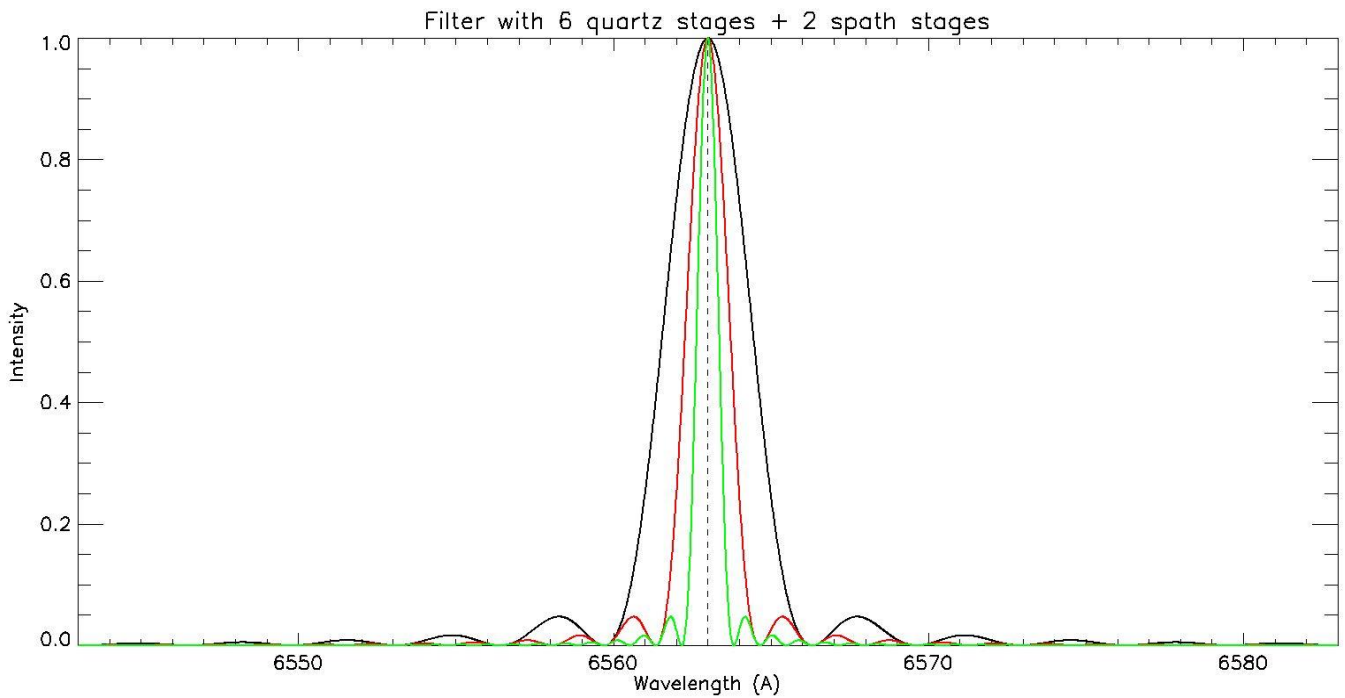

**Figure 15** : *The transmission of the Lyot chromospheric Hα filter. The original filter was in quartz (black line, 3.0 Å FWHM) and a spath stage was incorporated to reduce the bandpass to 1.5 Å FWHM (red curve). Later, a second spath was added to finally get 0.75 Å FWHM (green curve). Courtesy Paris observatory.*

Three filters made by OPL are preserved at Meudon (figure 16), together with pieces of the original Lyot filters. The three filters were checked in 2006 at the 14 m spectrograph of the Meudon Solar Tower, which allowed to visualize their bandpass.

Filter N° 2, temperature = 40.8° C, 0.75 Å FWHM

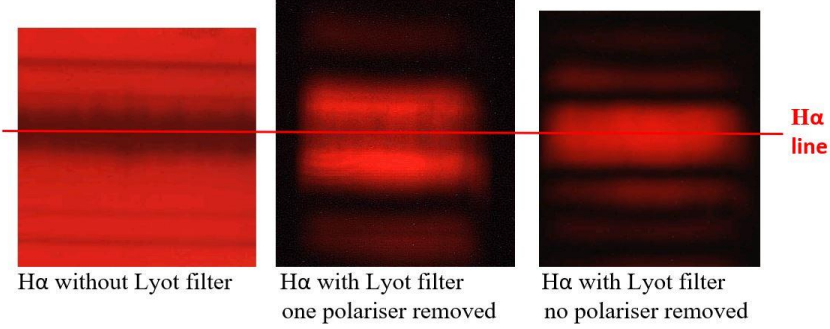

H$\alpha$ without Lyot filter | H$\alpha$ with Lyot filter one polariser removed | H$\alpha$ with Lyot filter no polariser removed

**H$\alpha$ line**

Filter N° 3, temperature = 27.5° C, 0.75 Å FWHM

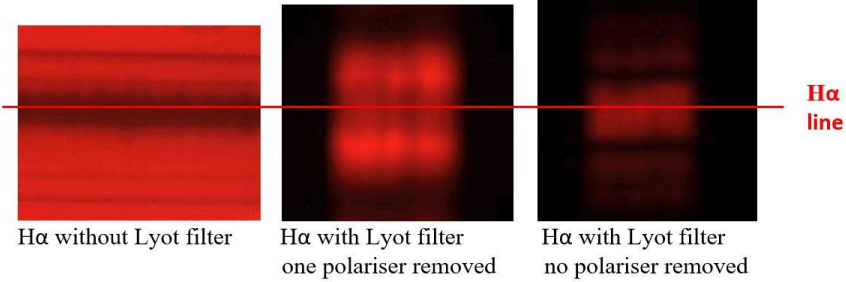

H$\alpha$ without Lyot filter | H$\alpha$ with Lyot filter one polariser removed | H$\alpha$ with Lyot filter no polariser removed

**H$\alpha$ line**

Filter N° 4, temperature = 44.0° C, 0.75 Å FWHM (one polarizer lacking)

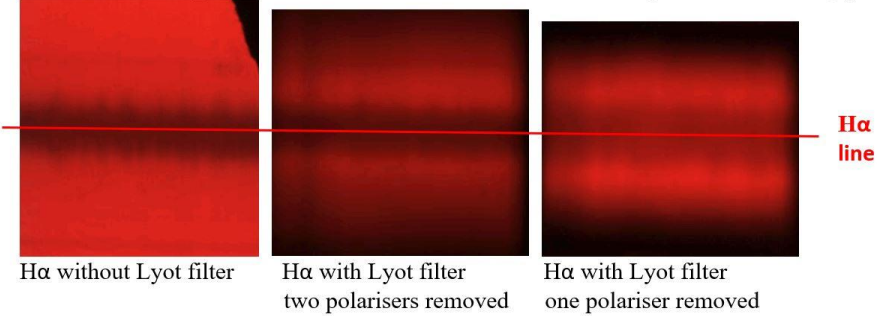

H$\alpha$ without Lyot filter | H$\alpha$ with Lyot filter two polarisers removed | H$\alpha$ with Lyot filter one polariser removed

**H$\alpha$ line**

**Figure 16** : *Optical tests of the three OPL filters conserved in Meudon. Left: the H$\alpha$ line observed at the spectrograph of Meudon Solar Tower. Centre: the H$\alpha$ line through the filters with one polarizer removed. Right: the same with all polarizers (except filter 4). Courtesy Paris observatory.*

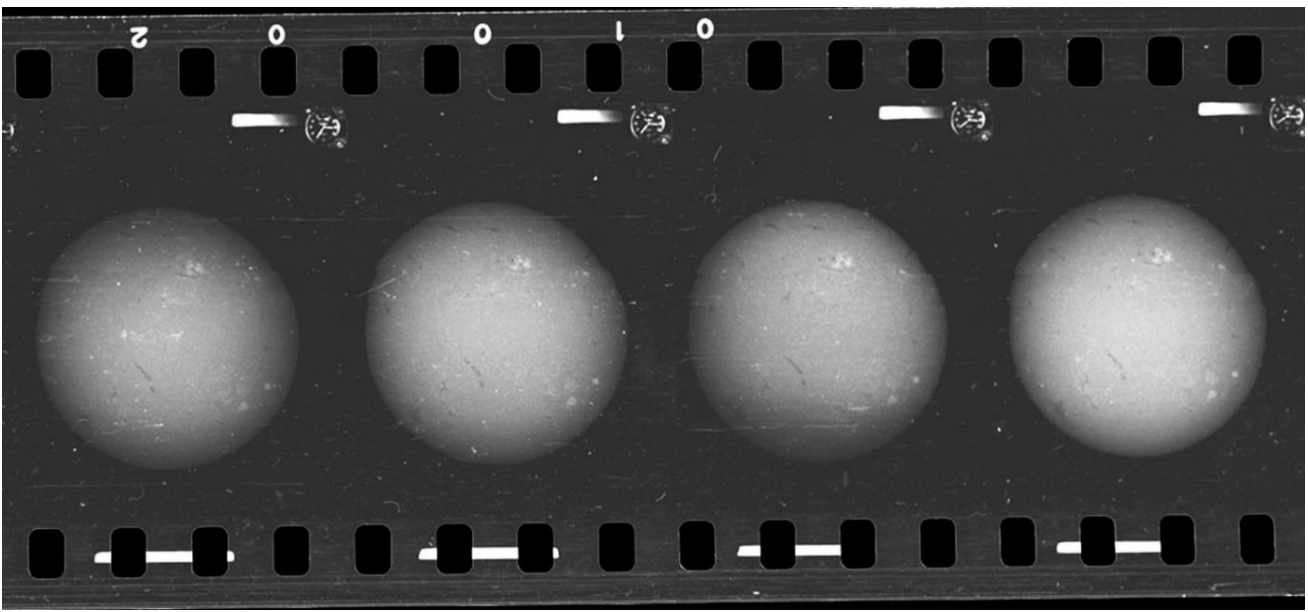

**Figure 17** : *H$\alpha$ images of the first Meudon heliograph (0.75 Å FWHM, 15 mm). Courtesy Paris observatory.*

Figure 17 shows typical frames of the 35 mm films produced by the first heliograph at Meudon (a part of the first film, June 1956). Movie 1 presents a short extraction of the 35 mm movie n°19 of 1957, as an example. Each movie is made of 1850 monochromatic images of the full Sun. From time to time, a long exposure picture is made for prominences at the limb (in that case, the disk is burned).

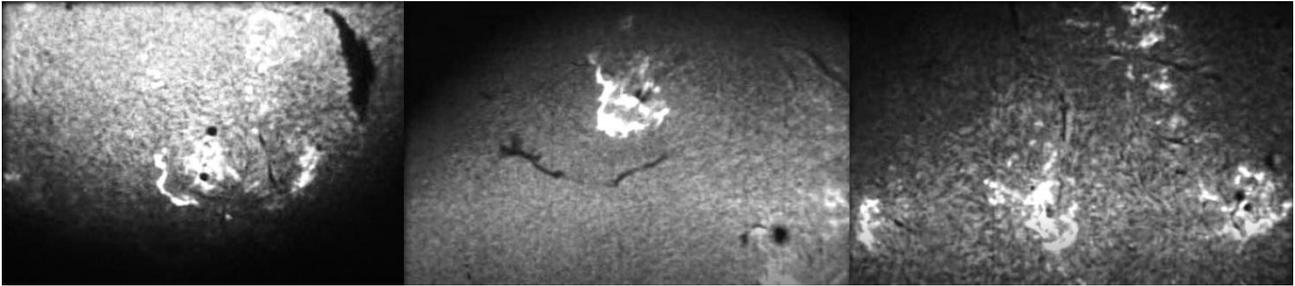

**Figure 18** : *Hα flares of 1957 and 1958 observed at Meudon. Courtesy Paris observatory.*

Figure 18 displays remarkable events of solar activity extracted from movie 2, which presents fast evolving active regions and flares observed in 1957 and 1958, during the rising phase of cycle n°19, in the frame of the International Geophysical Year (IGY1957).

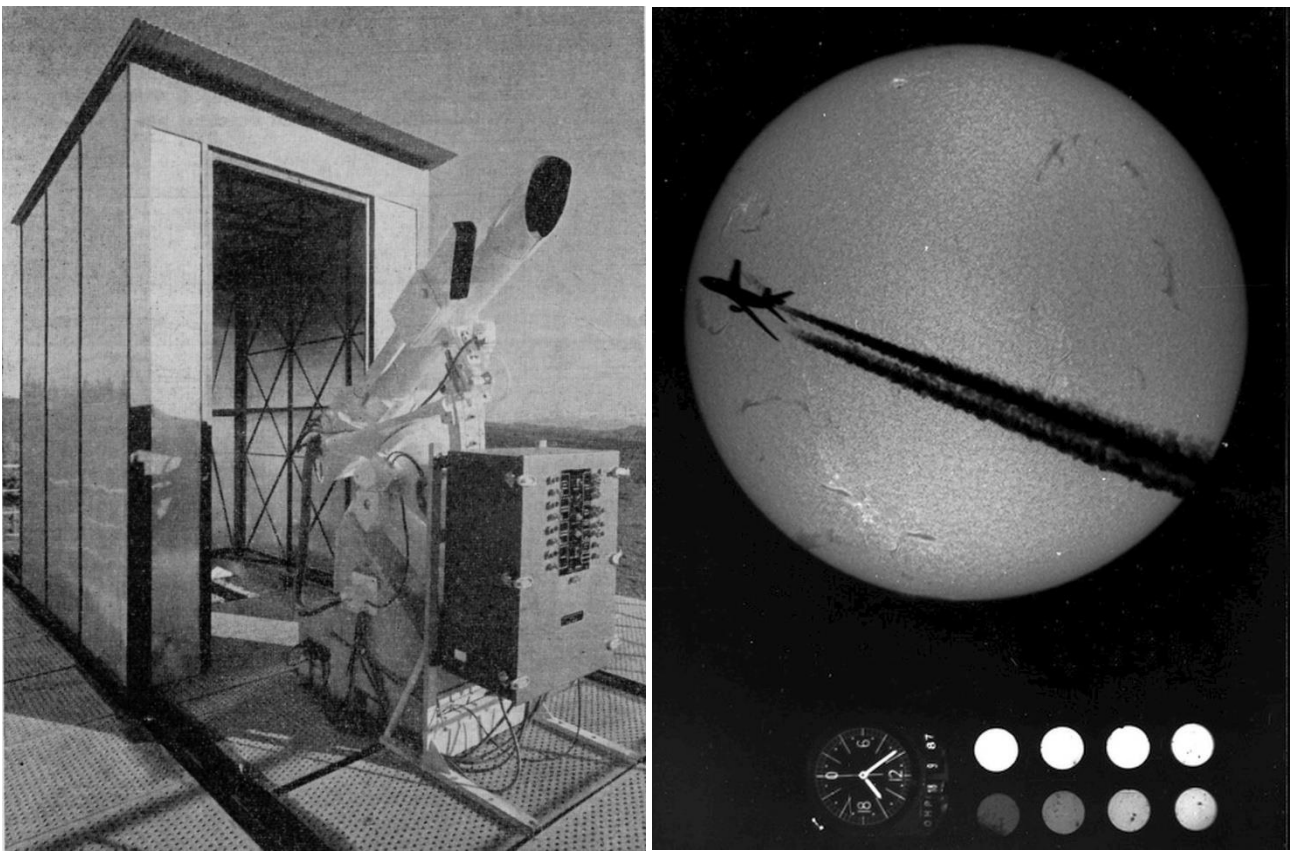

**Figure 19** : *The SECASI heliograph at OHP (left) and one of the Hα images of 18 September 1987 (right). Courtesy OHP and Paris observatory.*

At OHP, the SECASI heliograph (figure 19) with the OPL filter started observations in 1958, which lasted until the retirement of the observer in 1994. 1200 films and 2150000 images were produced by this instrument. The filter was periodically revisited in Meudon.

Movie 3 shows a part of the first film taken in 1958 with the OHP heliograph.

## 3 – CAPABILITIES OF MEUDON TUNABLE FILTERS (THE "VARIABLE WAVELENGTH")

In 1965, a new heliograph using a tunable filter, allowing to observe the Hα core and wings (± 0.75 Å), was installed at Meudon, but only for a part of the Sun, as the diameter of the solar image was 45 mm (twice the maximum size acceptable for 35 mm films). This instrument was designed for the survey of dynamic events, such as flares and filament eruptions (Michard, 1965). However, the full disk routine (figure 20) continued to run with an image reduced from 15 mm to 11 mm. The FWHM of both filters was 0.75 Å.

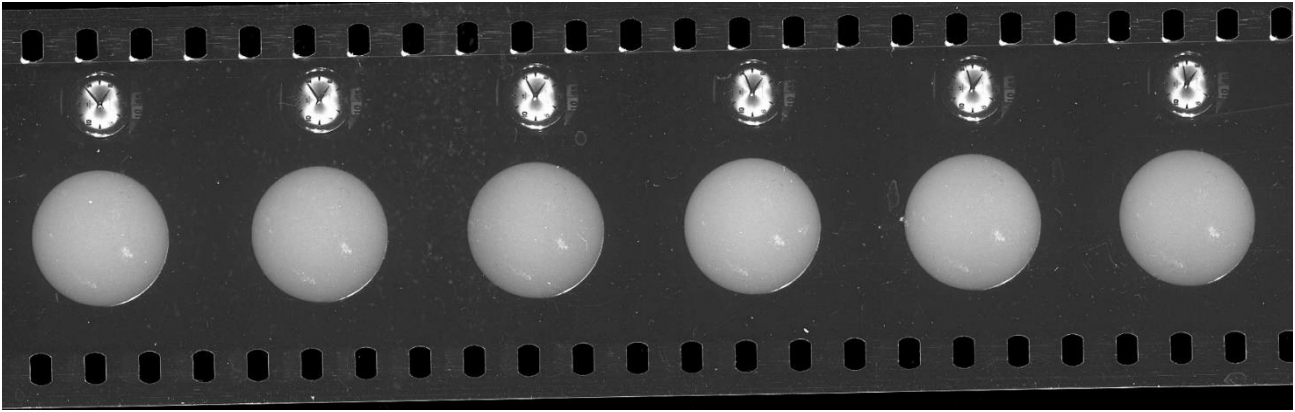

**Figure 20** : *The full disk Hα patrol after 1965 (11 mm diameter). Courtesy Paris observatory.*

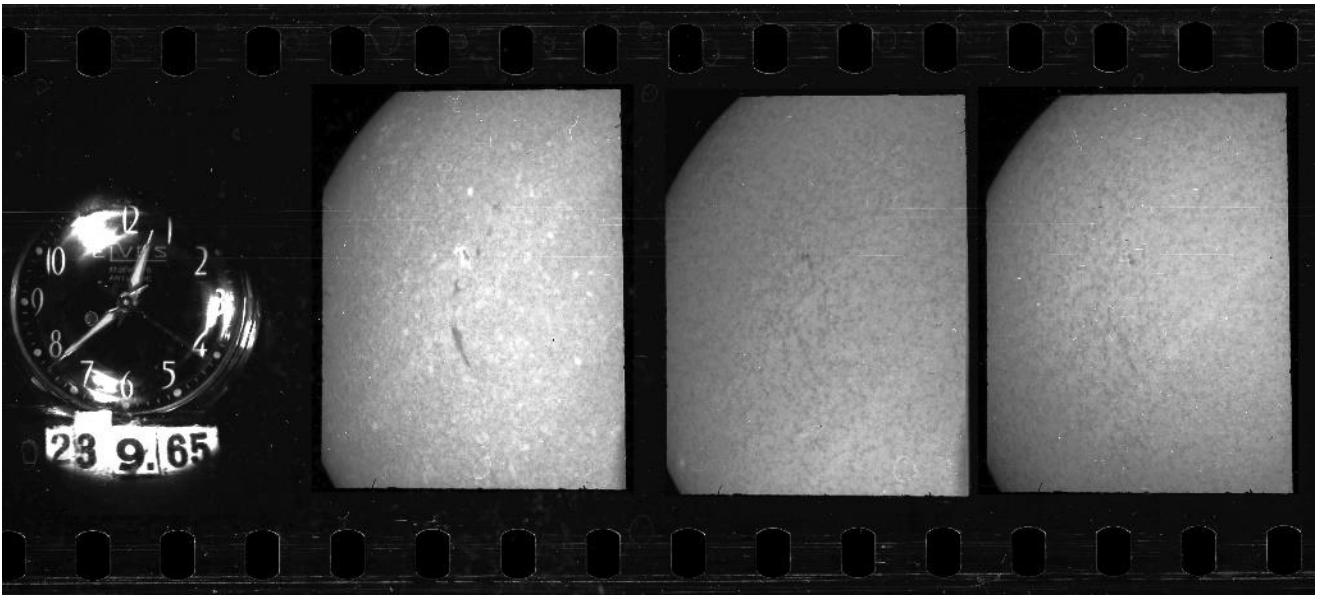

**Figure 21** : *The variable wavelength heliograph after 1965 (here 23 September 1965) with Hα centre, red and blue wings. The diameter of the Sun is 45 mm, so that a region is recorded on the 35 mm film.* 0.75 Å *FWHM. Courtesy Paris observatory.*

The three wavelength patrol of figure 21 is at the origin of many publications concerning dynamic phenomena; for instance, Banos (1967) studied an Hα flare with its counterparts in terms of radioelectric burts and magnetic fields; Mouradian *et al* (1983) demonstrated that a complex flare is composed of many elementary eruptive processes, involving both cold surging arches and hot flaring arches expanding in the solar atmosphere.

Movie 4 presents an extract of film n°44 (4 July 1970) taken with the "variable wavelength" heliograph of Meudon, concerning an active region (sunspot, filament). The three successive wavelengths are Hα centre and Hα ± 0.75 Å (blue and red wings). The observation of line wings provides a qualitative information about Dopplershihts (upward or downward motions, with the correspondance 0.75 Å = 34 km/s) and allows to detect fast motions, such as ejections or "disparitions brusques" of filaments.

The most powerful filter used at Meudon worked between 1985 and 2004. It was composed of two interlaced filters located inside an afocal optical design (figure 22). The refractor was open at F/15 (2.25 m focal length, 15 cm aperture for O1). The afocal system had a magnification 1.0 for 35 mm films (360 mm focal length for O2 and O3, providing a solar diameter of 21 mm), but it was reduced to 0.4 for the CCD after 1999 (140 mm focal length for O3, KAF1600 sensor, 1536 x 1024 pixels of 9 µ, solar diameter of 8.5 mm). There was also an ocular for visual inspection.

The filter (figures 23 and 24) was a tunable filter with 5 rotating stages, 0.5 Å FWHM, plus a prefilter (an interference filter of 3.5 Å FWHM) to isolate the Hα peak (Demarcq *et al,* 1985). It was in fact composed of two wide field and interlaced filters of spath, the first one with 3 stages (thickness e = 4.2 mm, 2 e, 4 e), and the second one with two stages (thickness d = 3/2 e = 6.3 mm, 2 d = 3 e). The tuning was obtained by rotating the stages (with respect to the fixed direction of the entrance polarizer) by α (thickness e), 2.5 α (thickness d = 1.5 e), 4.5 α (thickness 2 e), 7.5 α (thickness 2 d = 3 e) and 11.5 α (thickness 4 e). The distance between peaks was 11 Å, this is the reason why a narrow interference filter was needed to select the Hα peak. The Lyot filter and prefilter were both maintained at constant temperature. Their combined transmittance is reported in figure 25.

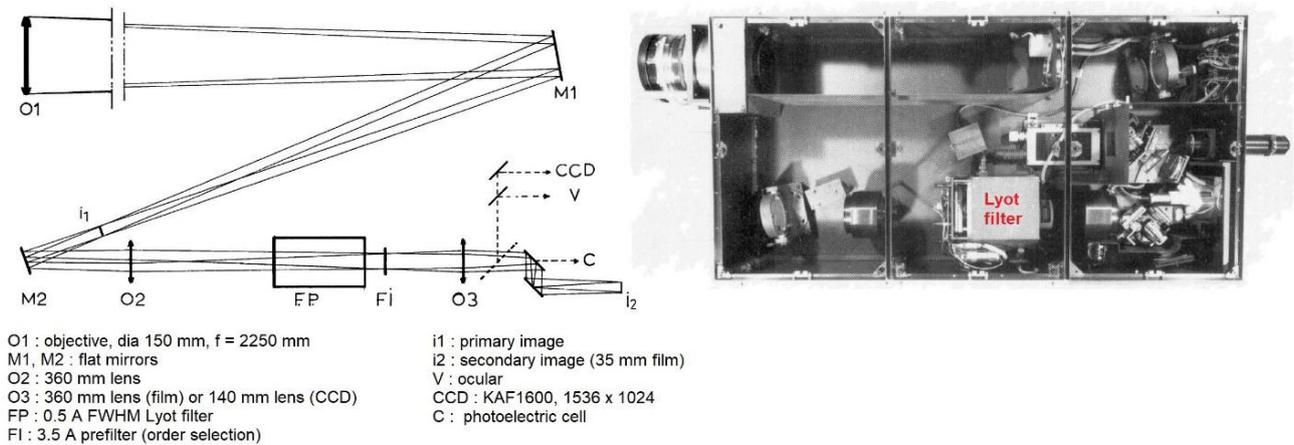

O1 : objective, dia 150 mm, f = 2250 mm
M1, M2 : flat mirrors
O2 : 360 mm lens
O3 : 360 mm lens (film) or 140 mm lens (CCD)
FP : 0.5 A FWHM Lyot filter
FI : 3.5 A prefilter (order selection)

i1 : primary image
i2 : secondary image (35 mm film)
V : ocular
CCD : KAF1600, 1536 x 1024
C : photoelectric cell

**Figure 22** : *Meudon heliograph with the Lyot tunable filter (0.5 Å FWHM). After Demarcq et al (1985) and courtesy Paris observatory.*

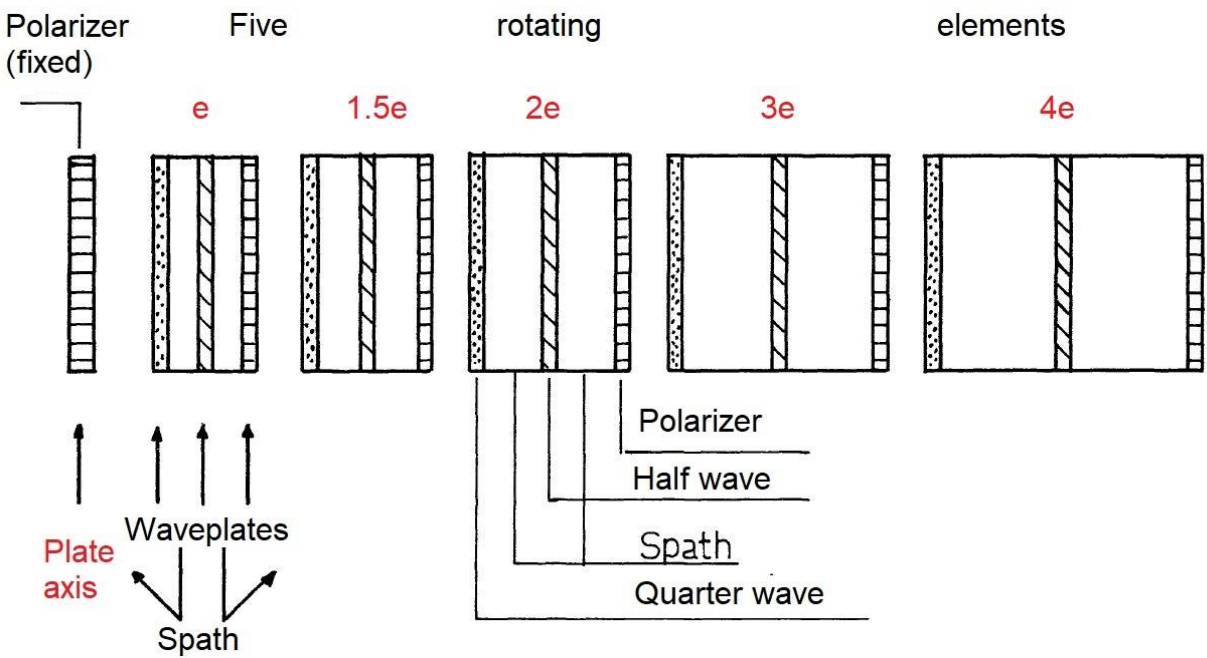

**Figure 23** : *Optical elements of the 5 stages Meudon tunable filter (0.5 Å FWHM, 1985), made of two interlaced filters (e, 2 e, 4 e) and (1.5 e, 3 e), with e = 4.2 mm (spath). Each stage rotates in an optical oil bath and is wide field. After Demarcq et al (1985).*

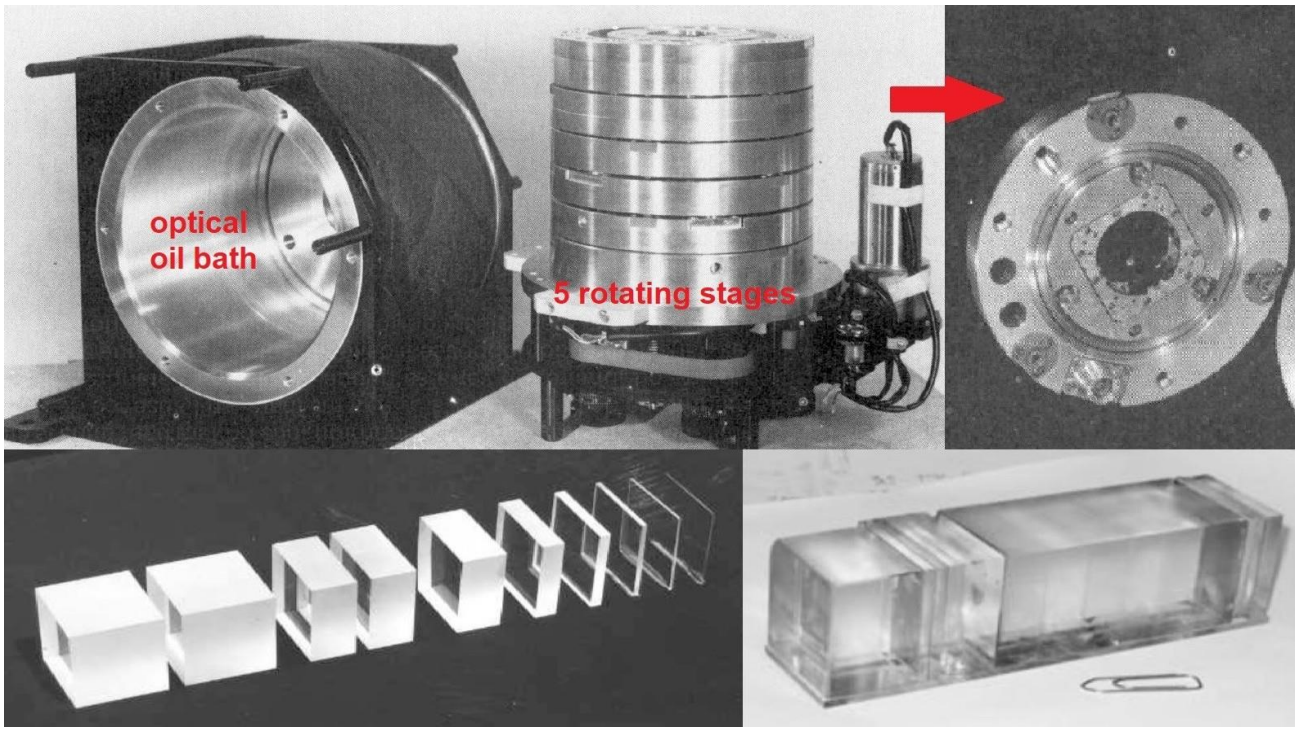

**Figure 24** : *Top: the Meudon tunable filter (1985) with 5 rotating stages. Bottom: for comparison, the classical assembly of glued optical elements of non tunable filters. Courtesy Paris observatory.*

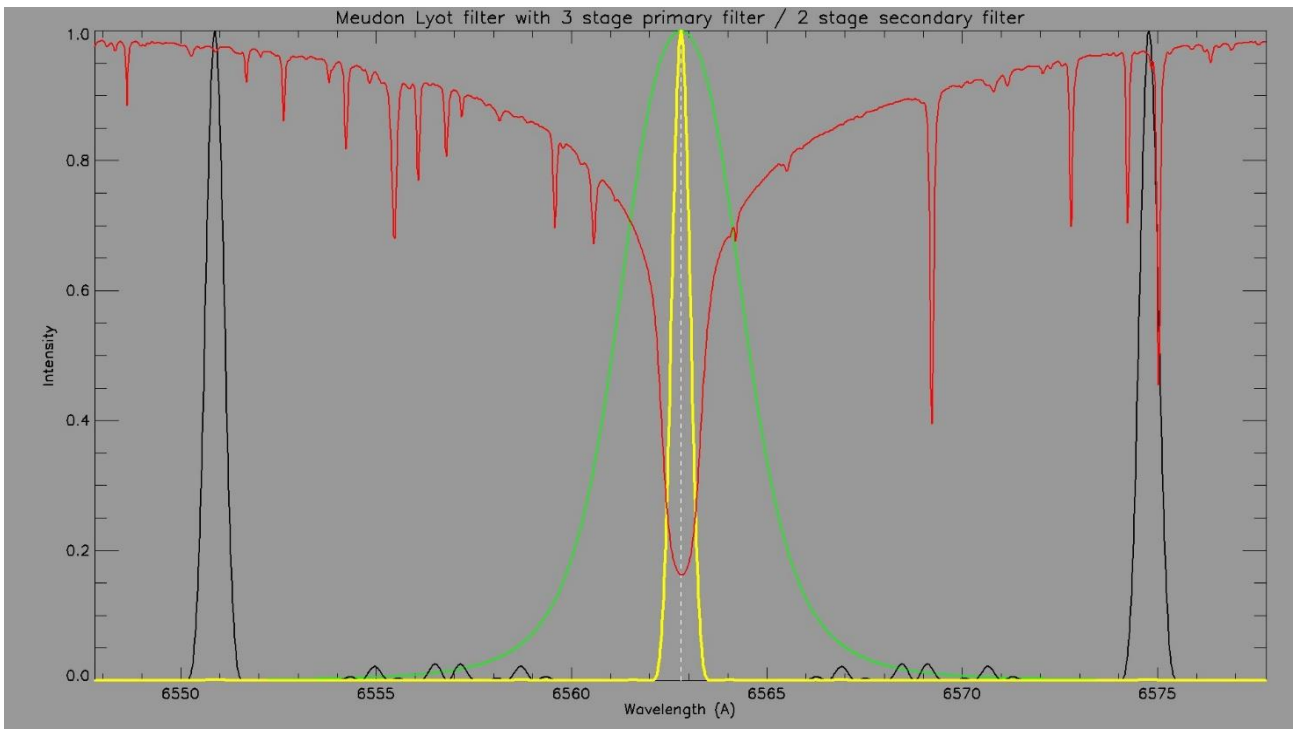

**Figure 25** : *Transmission of the 5 stages tunable Lyot filter. Black curve: the filter.* *Green curve: the prefilter.* *Yellow curve: the final transmittance.* *Red curve: the Hα line. Courtesy Paris observatory.*

Figure 26 displays the transmittance of the prefilter and of the 5 stages Lyot filter as observed at the large 14 m spectrograph of Meudon Solar Tower.

Movie 5 shows the exploration of the Hα line in the range [-1.0 Å, +1.0 Å] obtained by the rotation of the five birefringent plates of the filter; the bandpass could move by steps of 0.1 Å.

Movie 6 is a simulation done from a data-cube of Meudon spectroheliograph (monochromatic images along the Hα profile with 0.155 Å resolution); the spectroscopic cube is multiplied by the wavelength transmittance of the filter (figure 25 and movie 5) and the transmission peak moves in the range [-1.0 Å, +1.0 Å] in order to explore Dopplershifts. The wavelength position of the peak is determined by the rotation of the five birefringent plates of the filter.

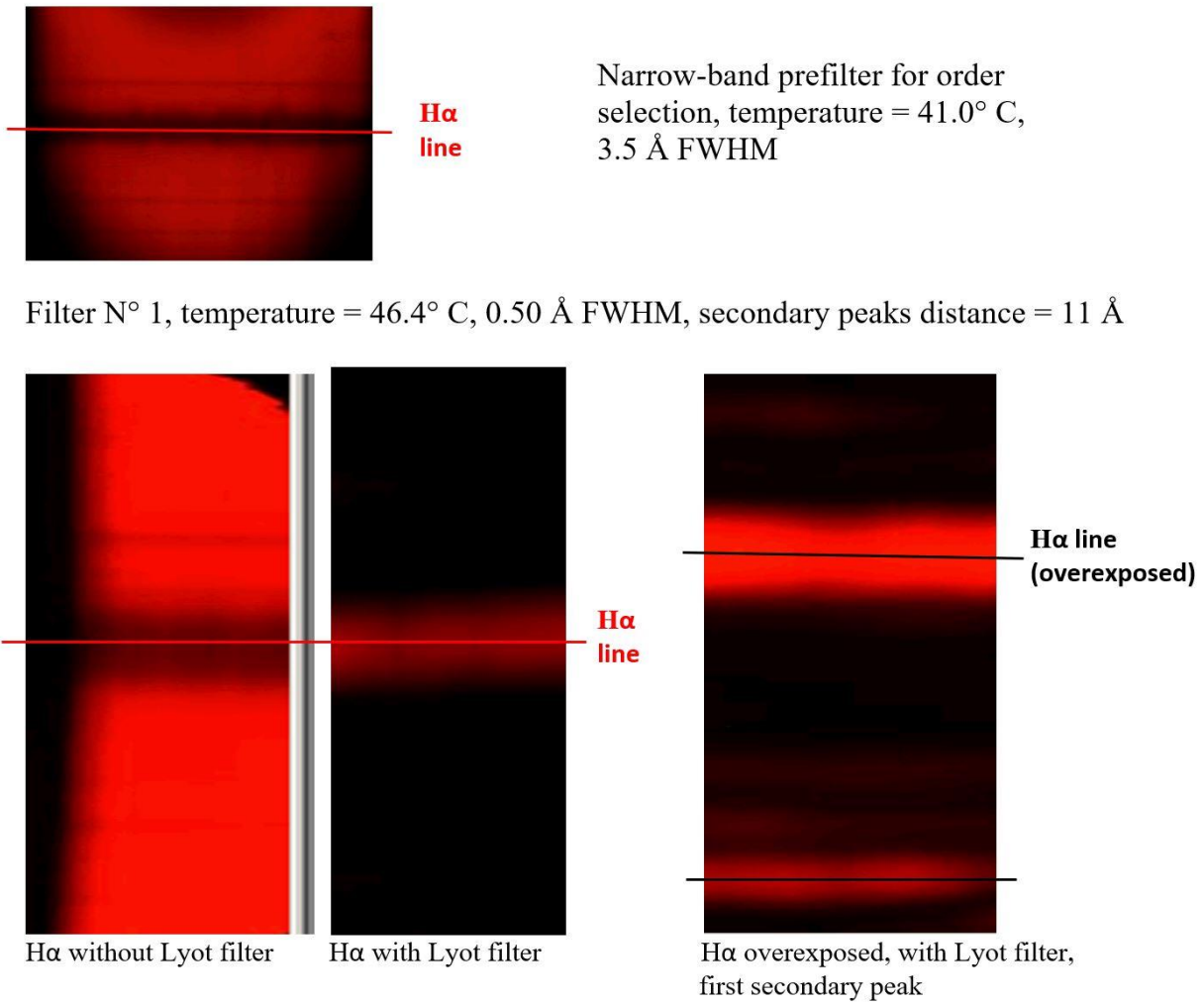

**Hα line**

Narrow-band prefilter for order selection, temperature = 41.0° C, 3.5 Å FWHM

Filter N° 1, temperature = 46.4° C, 0.50 Å FWHM, secondary peaks distance = 11 Å

**Hα line**

**Hα line (overexposed)**

Hα without Lyot filter    Hα with Lyot filter

Hα overexposed, with Lyot filter, first secondary peak

**Figure 26** : *Optical tests of the prefilter (top) and the 5 stages tunable Lyot filter (bottom). Left: the Hα line observed at the spectrograph of Meudon Solar Tower. Centre: the Hα line through the Lyot filter. Right: the same with overexposure. Courtesy Paris observatory.*

An example of observations with the tunable filter, on 02 June 1991, is given by figure 27.

Movie 7 shows an example of observations (2 May 1990, film n°23 of 1990) with three wavelengths positions, obtained by the rotation of the birefringent plates of the filter. The three successive wavelengths are Hα centre and Hα ± 0. 50 Å (blue and red wings). The observation of line wings provides informations about Dopplershihts (upward or downward motions, with the correspondance 0.50 Å = 23 km/s) and allows to detect fast motions, such as ejections or "disparitions brusques" of filaments.

Movie 8 shows another example of observations (28 October 2003, CCD camera) with three wavelengths positions, obtained by the rotation of the birefringent plates of the filter. Meudon heliograph observed the famous and huge X17 event, a "two ribbon" flare occurring near the disk centre around 11:00 TU, with impact at the Earth. At the end of the movie clip, the running differences in the three wavelengths positions reveal a fast moving Moreton wave (1000 km/s), particularly well visible in the line wings. The Moreton wave is probably the chromospheric counterpart of an MHD fast shock propagating in the low corona and compressing the chromosphere below. As suggested by the movie, the compression seems to be characterized by redshifts of the Hα line (chromospheric downward motion), while the relaxation of the chromosphere (after the passage of the coronal shock) could be accompanied by blueshifts.

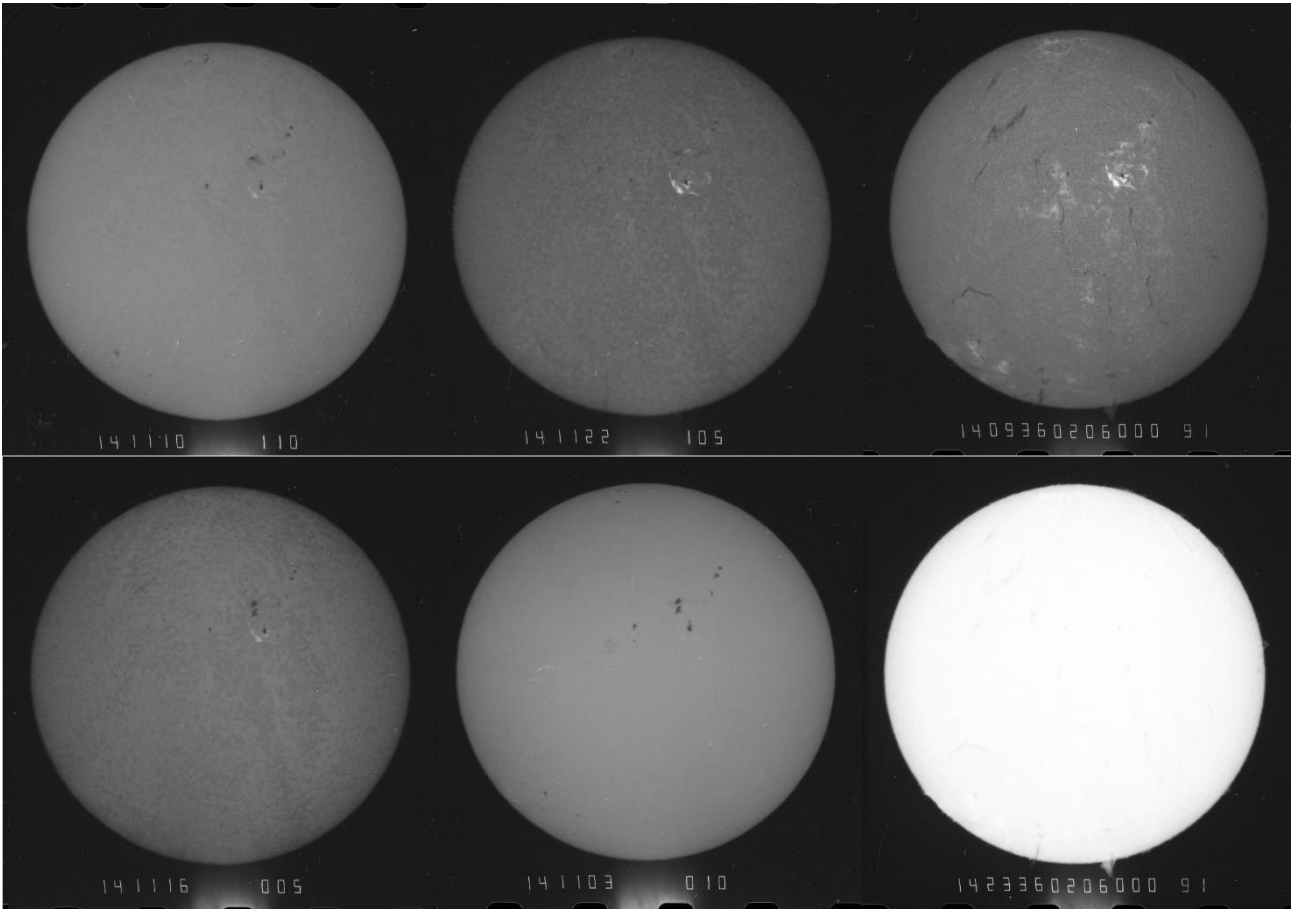

**Figure 27** : *An exemple of observations (02 June 1991) with the 5 stages tunable filter. Top: Hα – 1.0 Å, Hα – 0.5 Å, Hα centre. Bottom: Hα + 0.5 Å, Hα + 1.0 Å, Hα centre (overexposed for prominences), Courtesy Paris observatory.*

## 4 – OBSERVATIONS

Observations performed at Meudon and OHP between 1956 and 2004 are summarized by figure 28 and Table 1. The collection contains 3000 films of 1850 frames and 700000 CCD images for a total of more than 6 million monochromatic images.

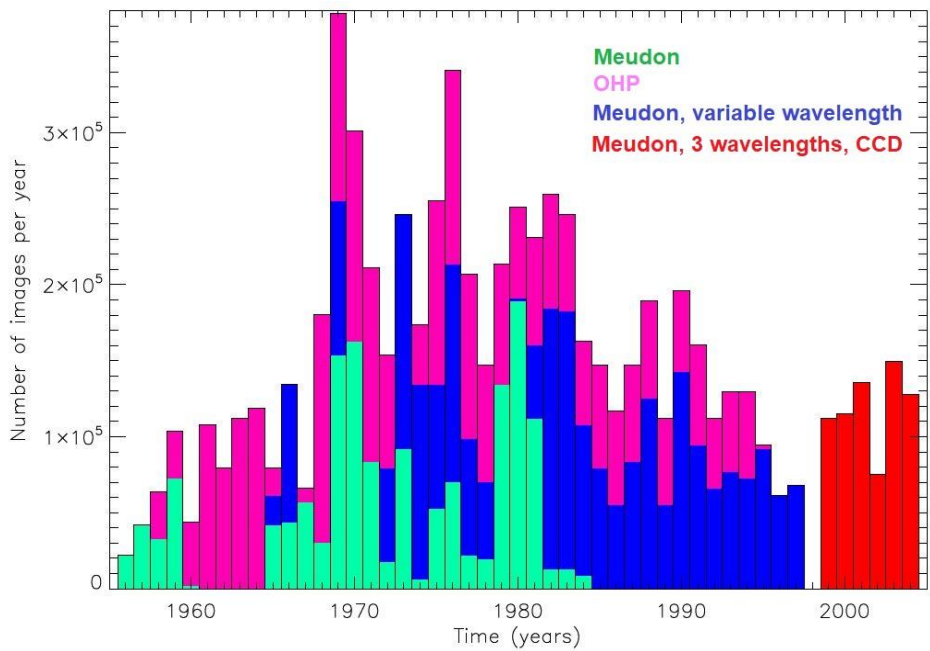

**Figure 28** :
*Observations of Meudon and OHP Hα heliographs. Courtesy Paris observatory.*

| Year | Meudon | Instrument | OHP | Instrument |
|---|---|---|---|---|
| 1956 | 10 FILMS | Sun diameter 15 mm<br>0.75 A FWHM | | |
| 1957 | 19 | | | |
| 1958 | 15 | | 14 | Sun diameter 15 mm<br>0.75 A FWHM |
| 1959 | 33 | | 14 | |
| 1960 | 1 | Interruption for new setup | 19 | |
| 1961 | 0 | | 49 | |
| 1962 | 0 | | 36 | |
| 1963 | 0 | | 51 | |
| 1964 | 0 | | 54 | |
| 1965 | 19 full sun<br>9 for regions, variable λ | 2 instruments, same mount<br>0.75 A FWHM<br>diameter 11 mm (full sun)<br>diameter 45 mm (regions) | 8 | |
| 1966 | 20 full sun<br>41 for regions, variable λ | | 0 | |
| 1967 | 26 | | 4 | |
| 1968 | 14 | | 68 | |
| 1969 | 70 full sun<br>46 for regions, variable λ | | 56 | |
| 1970 | 74 | | 63 | |
| 1971 | 38 | | 58 | |
| 1972 | 8 full sun<br>28 for regions, variable λ | | 34 | |
| 1973 | 42 full Sun<br>70 for regions, variable λ | | 0 | |
| 1974 | 3 full sun<br>58 for regions, variable λ | | 18 | |
| 1975 | 24 full Sun<br>37 for regions, variable λ | | 55 | |
| 1976 | 32 full Sun<br>65 for regions, variable λ | | 58 | |
| 1977 | 10 full Sun<br>35 for regions, variable λ | | 49 | |
| 1978 | 9 full Sun<br>23 for regions, variable λ | | 35 | |
| 1979 | 61 films soleil 11mm | | 36 | |
| 1980 | 86 full Sun<br>1 for regions, variable λ | | 27 | |
| 1981 | 51 full Sun | | 32 | |
| 1982 | 6 full Sun<br>78 for regions, variable λ | | 34 | |
| 1983 | 6 full Sun<br>77 for regions, variable λ | 2 instruments, same mount<br>0.75 A FWHM<br>diameter 16 mm (full sun)<br>diameter 35 mm (regions) | 29 | |
| 1984 | 4 full Sun<br>45 for regions, variable λ | | 25 | |
| 1985 | 36 full Sun, variable λ | Sun diameter 21 mm<br>0.5 A FWHM | 31 | |
| 1986 | 25 | | 28 | |
| 1987 | 38 | | 29 | |
| 1988 | 57 | | 29 | |
| 1989 | 25 | | 26 | |
| 1990 | 65 | | 24 | |
| 1991 | 43 | | 30 | |
| 1992 | 30 | | 21 | |
| 1993 | 35 | | 24 | |
| 1994 | 33 | | 26 | |
| 1995 | 42 | | 1 | End OHP |
| 1996 | 28 | | | |
| 1997 | 31 | | | |
| 1998 | 0 | | | |
| 1999 | Full Sun<br>110000 images | Sun diameter 8.5 mm<br>0.5 A FWHM<br>CCD KAF1400 | | |
| 2000 | 115000 images | | | |
| 2001 | 135000 images | | | |
| 2002 | 75000 images | | | |
| 2003 | 150000 images | | | |
| 2004 | 130000 images | End Meudon | | |

**Table 1** : Meudon and OHP observations from 1956 to 2004 (number of 45 m films, 1 film = 1850 images)

Catalogs of films and CCD images are available on line (see below). But data sets are not available on line; CCD images are stored on CD-ROM and DVD-ROM and can be requested via email to: observateurs.solaires@obspm.fr. Data recorded on films can also be requested but require a longer delay, because of the time necessary for the digitization (one film = 1850 eight bits TIF files).

## 5 – THE METEOSPACE PROJECT

The Hα Meudon routines were stopped in 2004 because of technical failures and lack of manpower. The international context also changed with the birth of networks such as GONG Hα (7 stations in good sites at various longitudes). It was decided to start a new project (MeteoSpace, figure 29) in collaboration with Côte d'Azur Observatory (OCA). The lack of observers and operators imposed the choice of fully automatic telescopes. The existence of other stations in good sites imposed also the choice of a sunnier site, such as the Calern plateau at 1270 m elevation, near Grasse and Nice (France). At last, the lost of knowledge of Lyot filters, due to many uncompensated retirements, together with a small budget, imposed finally the choice of commercial Fabry-Pérot filters such as the DayStar Quantum Pro series.

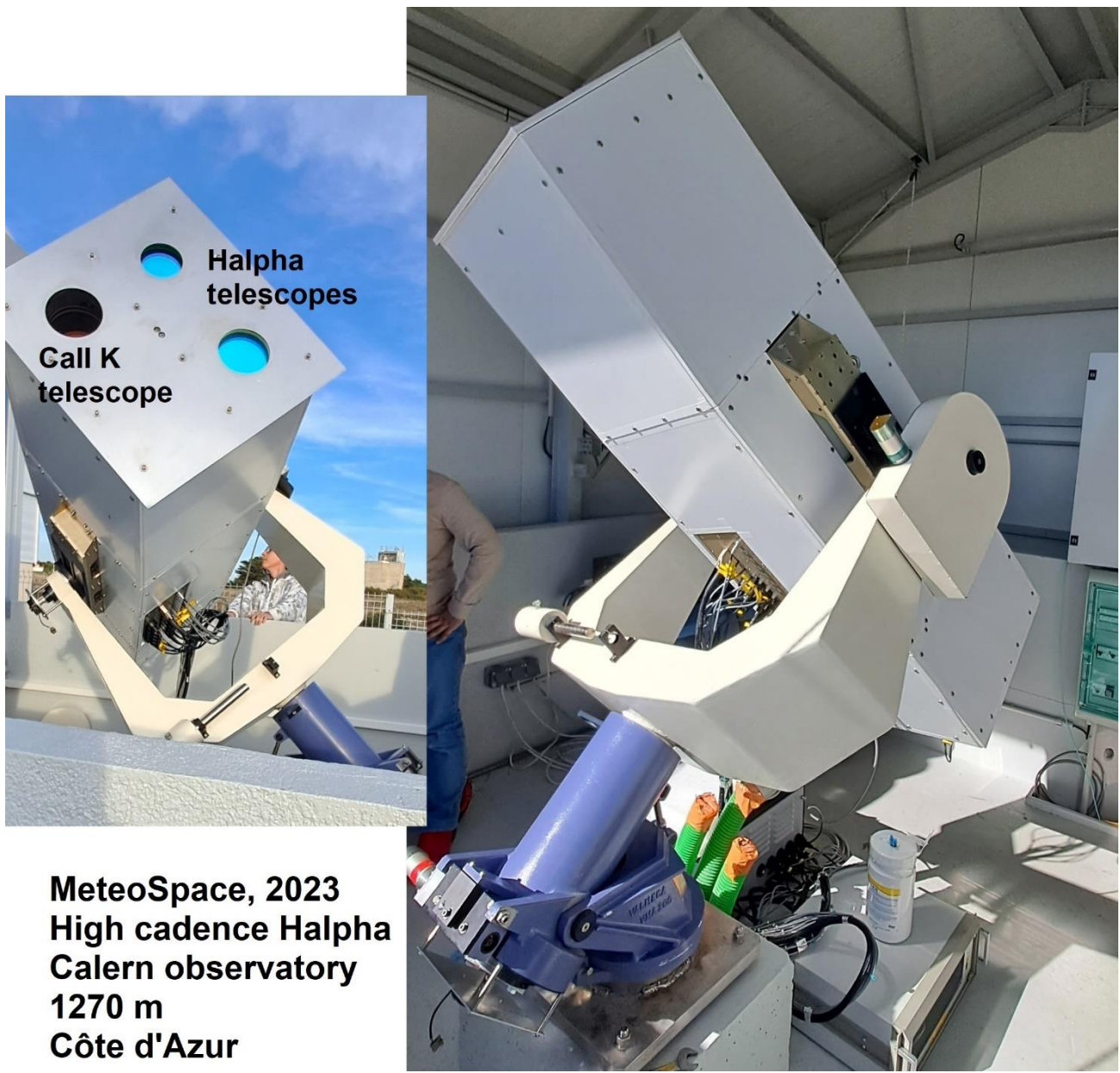

**Figure 29** : *The MeteoSpace telescopes in 2023 at Calern observatory. Courtesy OCA.*

The MeteoSpace instrument will be commissioned in the course of 2023 and is dedicated to observe fast events of solar activity during cycle 25, until 2030 (at least).

MeteoSpace is a fully automatic instrument providing real time JPEG images for quick look and FITS images for scientific use. The data base will be located in Nice, but the portail will be the solar Web site https://bass2000.obspm.fr located in Meudon. Data will be freely available to the international community.

MeteoSpace carries three telescopes, including Hα centre, Hα wing (red or blue) for the detection of Moreton waves (both 0.5 Å FWHM, figure 30), and a CaII K channel (1.5 Å FWHM) used as a proxy for magnetic fields. The temporal resolution is close to 10 s, in order to record the detailed evolution of dynamic events.

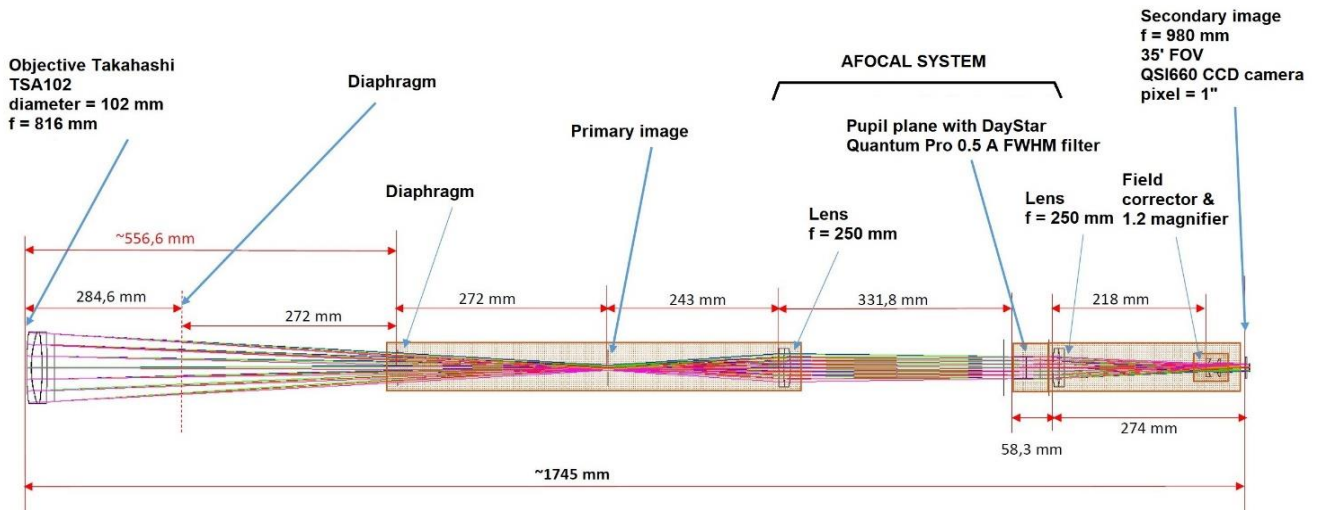

**Figure 30** : *The Hα telescopes of the MeteoSpace project. Courtesy Paris/OCA observatories.*

The Hα telescopes are open at F/10 and use an afocal system at F/30 including a 0.5 Å FWHM DayStar Quantum Pro filter with Lorenztian shape, so that images will have a reduced contrast in comparison to the Lyot filter technology with similar bandpass. The equivalent focal length is 0.983 m. The CCD cameras use Sony interline sensors with 4.5 µ pixels. The spatial sampling is about 1". The entire instrument is kept at constant temperature with active heating at 28°C and passive cooling. Details can be found in Malherbe *et al* (2019, 2022). MeteoSpace is located inside a motorized rolling house with curtains. Operations are fully automated and are driven by a supervisor computer, which opens the dome, catches the Sun, starts data acquisition, processes images and executes file transfers to the Nice data centre. Many meteorological and environment sensors are checked permanently in order to detect clouds and wind. In case of risk, the computer decides to interrupt observations and closes the dome. First observations should start in Summer 2023 in order to observe the rising phase of cycle 25, and major flares at the maximum and after.

## 6 – CONCLUSION

We described in this paper the evolution of the successive generations of monochromatic heliographs running at Meudon and OHP observatories between 1956 and 2004, in terms of optical characteristics and capabilities. These instruments are at the origin of more than 6 million pictures, which are unfortunately not on-line, contrarily to the catalogs, but which can be put on-line upon request. The survey of Hα solar flares and associated events (Moreton waves, "disparitions brusques" of filaments, eruptive prominences, mass ejections) is definitely stopped at Meudon (apart the spectroheliograph with a few images per day), but it will recover soon at Calern observatory with the MeteoSpace project designed for observations of the solar cycle 25, at high temporal cadence (5 images/minute).

## 7 - REFERENCES

## 8 – NOTES

**[1] The solar atmosphere**. It is composed of three layers: (1) the photosphere (the visible surface, temperature decreasing from 6000 K to 4500 K in 300 km) with dark sunspots and bright faculae around, (2) the chromosphere above (temperature increasing from 4500 K to 8000 K in 2000 km) with dark filaments and bright plages (corresponding to faculae in the photosphere), (3) the ionized and hot corona (2 million K). The corona extends at million kilometres and gives rise to the solar wind (charged particles). The chromosphere is the source of solar activity (flares, ejections) and requires monochromatic observations to reveal the structures via absorption lines (such as Hα) of the visible spectrum. The solar atmosphere follows a 11-years activity cycle and a 22-years magnetic cycle. Flares and ejections occur in active regions a few years around the solar maximum; the maximum of the current cycle (number 25) is forecasted for 2025.

**[2] The solar structures**. Dark sunspots are regions of intense magnetic fields (0.1 T - 0.3 T). Bright faculae or plages form, together with sunspots, active regions, and exhibit smaller magnetic fields (0.01 T - 0.05 T). Dark filaments, also called prominences when seen at the limb, are thin and high structures (50000 km) of dense material suspended in the corona by weak magnetic fields (0.001 T). Flares occur in zones of unstable fields; reconnections drive fast events and convert magnetic energy into kinetic energy (ejections), radiation (X-rays) and heat (brightenings).

**[3] The Doppler effect**. When a light source moves in the observer's direction, the spectral lines are shifted towards shorter wavelengths (blueshift). On the contrary, if the source moves in the opposite direction, lines are shifted towards longer wavelengths (redshift). The Dopplershift w(Å) is a wavelength shift, proportional to

the projection V of the velocity vector along the line-of-sight (also called radial velocity, positive towards the observer in solar physics): w = - λ (V/C), where λ is the line wavelength (Å) and C the light speed ($3 \times 10^5$ km/s).

## 9 - ON-LINE MATERIAL (CATALOGS, MPEG4 MOVIES, FIGURES)

### a) CATALOGS OF OBSERVATIONS

Catalog of OHP observations (35 mm films, 1958-1996)

https://drive.google.com/file/d/1Cszx3orKUn2pNHuz_IwGXTcMkVcNHSB8/view?usp=share_link

Catalog of MEUDON observations (35 mm films, 1956-1997)

https://drive.google.com/file/d/1mRLOgle_I7X1FwwUAiUPhRkscvhNxg2c/view?usp=share_link

Catalog of MEUDON observations (CCD, 1999-2004)

https://drive.google.com/file/d/1mKjZtXYchipEtB6XmGMwZvJjTsAOvJgp/view?usp=share_link

### b) FIGURES, MPEG4 MOVIES

**Figures :**

https://drive.google.com/drive/folders/1NDJ84VF8WoOGFqZwcDrsDaEwmZ5qioU_?usp=share_link

**MPEG4 video movies (courtesy Paris Observatory) :**

Movie 1: Meudon, part of the 35 mm film n°19, 1957. Hα centre. Lyot filter, 0.75 Å FWHM.
https://drive.google.com/file/d/1xHkgUuJHCdlxzI45sko4gmKASMl2nOz1/view?usp=share_link

Movie 2: Examples of solar activity seen in Hα centre and recorded during the International Geophysical Year (IGY 1957). Lyot filter, 0.75 Å FWHM. Meudon heliograph. 35 mm film.
https://drive.google.com/file/d/1EZw4X1o2P8qCQTChHQu0MHSmSJC7t5CJ/view?usp=share_link

Movie 3: Part of the first film of OHP, n°1, 1958. Hα centre. Lyot filter, 0.75 Å FWHM.
https://drive.google.com/file/d/11mjxQj-NJTC3ZqoRwjfiEGCGQUYyT5CWf/view?usp=share_link

Movie 4: Meudon, first "variable wavelength" heliograph for regions. 1965 version. Extract of the 35 mm film n°44, 1970. Solar active region (part of the Sun). Hα centre and Hα ± 0.75 Å (blue and red wings). Tunable Lyot filter, 0.75 Å FWHM.
https://drive.google.com/file/d/1UMwKiKXdrKH2LG8-YvA3fuhH4_0XNWwl/view?usp=share_link

Movie 5: Transmission of the Meudon Hα tunable Lyot filter, 1985 version, 0.50 Å FWHM. Wavelength scan by rotating the five birefringent plates. In abscissa: the wavelength (Å). Red curve: the line profile of Hα. Solid line: the transmission of the filter (3 peaks are visible). Green curve: the prefilter transmission (3.50 Å FWHM). Yellow curve: the product of the transmission curves of the prefilter and of the Lyot filter showing a single peak. The filter can be tuned between Hα – 1.0 Å and Hα + 1.0 Å.
https://drive.google.com/file/d/1hlsvdyrAZSyVbKA4S22CCm4mjvQRuMm8/view?usp=share_link

Movie 6: Meudon Hα tunable Lyot filter, 1985 version, 0.50 Å FWHM. Simulation of the tuning between Hα – 1.0 Å and Hα + 1.0 Å using a data-cube of Meudon Spectroheliograph (0.155 Å resolution). The data-cube is multiplicated (in wavelength) by the transmission of the filter (movie 5) and the result is integrated over wavelengths to form a monochromatic image centred on the filter bandpass.
https://drive.google.com/file/d/1bU31E8X8_by40zodFCZ20EG2xZM5lpq4/view?usp=share_link

Movie 7: Meudon Hα tunable Lyot filter, 1985 version, 0.50 Å FWHM. An example of observations with three wavelengths (2 May 1990, part of the 35 mm film n°23, 1990), Hα centre and Hα ± 0.50 Å (blue and red wings) plus (from time to time) an overexposed picture for prominences at the limb.
https://drive.google.com/file/d/1i1-cF6R-T-rWD_njJ16rpFl8gGh_rcoe/view?usp=share_link

Movie 8: Meudon Hα tunable Lyot filter, 1985 version, 0.50 Å FWHM. An example of observations with three wavelengths (28 May 1990). CCD version after 1999, Hα centre and Hα ± 0.50 Å (blue and red wings). A major flare (X17) occurs at 11:00 UT. At the end of the movie, running differences of Hα centre and Hα ± 0.50 Å at flare time showing a fast Moreton wave, well visible in the two wings of the line (bottom row).
https://drive.google.com/file/d/113OzybtO5WfYstvZBcx92P_UNVREkJa0/view?usp=share_link

## 10 - ACKNOWLEDGEMENTS

The author thanks I. Bualé and F. Cornu for images and archives of the Meudon solar collection, and for the catalogs of Meudon and OHP observations. He is indebted to Ch. Coutard for testing of the Lyot filters. He acknowledges Th. Corbard, G. Barbary, C. Collin, and F. Morand for the MeteoSpace project.

## 11 - THE AUTHOR

Dr Jean-Marie Malherbe, born in 1956, is astronomer at Paris-Meudon observatory. He got the degrees of "*Docteur en astrophysique*" in 1983 and "*Docteur ès Sciences*" in 1987. He used the spectrographs of the Meudon Solar Tower, the Pic du Midi Turret Dome, the German Vacuum Tower Telescope, THEMIS (Tenerife) and developed polarimeters. He also used the space-born instruments HINODE (JAXA), SDO and IRIS (NASA). He is responsible of the Meudon spectroheliograph since 1996.